\documentclass[a4paper,11pt]{article}
\usepackage{jheppub}
\usepackage[utf8]{inputenc}
\usepackage{amsthm,bm,mathtools,mathrsfs}
\usepackage{microtype}
\usepackage{booktabs,array}
\usepackage{enumitem}

\hypersetup{
  pdftitle={Generalized Coordinate Bethe Ansatz and Quantum Integrability of the Time-Dependent Massive Thirring Model},
  pdfauthor={Parameshwar R. Pasnoori}
}

\allowdisplaybreaks[2]
\numberwithin{equation}{section}

\newcommand{\ii}{\mathrm{i}}
\newcommand{\dd}{\mathrm{d}}
\newcommand{\mb}{m_{\mathrm b}}

\newcommand{\bbet}{\boldsymbol\beta}

\newcommand{\bx}{\boldsymbol x}
\newcommand{\bn}{\boldsymbol n}
\newcommand{\be}{\boldsymbol e}

\newcommand{\ket}[1]{\left|#1\right\rangle}
\newcommand{\Smu}{S_\mu}

\makeatletter
\gdef\@fpheader{}
\makeatother

\title{\boldmath Generalized Coordinate Bethe Ansatz and Quantum Integrability of the Time-Dependent Massive Thirring Model}

\author[a]{Parameshwar R. Pasnoori}
\affiliation[a]{Mani L. Bhaumik Institute for Theoretical Physics, Department of Physics and Astronomy, University of California, Los Angeles, CA 90095, USA}
\emailAdd{parmesh12@g.ucla.edu}

\abstract{
We develop a generalized coordinate Bethe-ansatz framework for the massive Thirring model with an explicitly time-dependent fermion mass. The construction is fundamentally different from the previously developed generalized Bethe ansatz for chiral quantum field theories. There, the exact wavefunction is associated with transport of particles in the configuration space involving operators acting in the internal spin spaces of the particles, which leads to quantum
Knizhnik-Zamolodchikov equations.  In the massive theory considered
here, the nontrivial structure instead appears directly in coordinate
space and in the physical rapidities. The central new feature is a \emph{rapidity lattice}.  Because the
time-dependent mass couples the right- and left-moving components
already in the one-particle sector, a single rapidity-labelled
one-particle wavefunction is not closed under spatial winding.
Periodicity of the complete wavefunction therefore requires all rapidity sectors generated by the finite shift $\beta\to\beta-p$, forming a discrete rapidity lattice. The
many-body wavefunction is constructed directly on these rapidity
lattices, while particle exchange determines the relative amplitudes
between their different sectors through scalar analytic difference
equations of Ruijsenaars type. Requiring the coordinate construction to be consistent also constrains
the time dependence itself and selects an exponential evolution of the
bare mass.  In the static limit the rapidity-lattice construction
reduces to the conventional massive-Thirring Bethe equations.
The result provides a new coordinate route to exact real-time dynamics
in an interacting relativistic quantum field theory.  Through the
massive-Thirring/sine-Gordon correspondence, it also gives a quantum
framework for comparing the generalized coordinate Bethe ansatz with
time-dependent Lax integrability and the time-dependent integrability-renormalization
group flow correspondence.
}


\begin{document}
\maketitle
\flushbottom

\section{Introduction}
\label{sec:intro}

Quantum integrability provides one of the few settings in which
strongly interacting many-body systems can be solved exactly.  In one
spatial dimension, the Bethe ansatz organizes the many-body problem in
terms of factorized two-particle scattering.  In its coordinate
formulation, configuration space is divided into sectors corresponding
to different orderings of the particles, and amplitudes in neighboring
sectors are related by two-particle $S$-matrices.  Consistency of
different sequences of exchanges is encoded in the Yang-Baxter
equation, while imposing periodic boundary conditions gives the Bethe
equations that quantize the rapidities.  This framework underlies the
exact solution of a broad class of quantum chains, interacting
one-dimensional systems and relativistic quantum field theories
\cite{Bethe1931,Yang1967,BergknoffThacker1979,Thacker1981}.

The same integrable structure admits an algebraic formulation through the quantum inverse-scattering method (QISM) and the algebraic Bethe ansatz \cite{SklyaninQISM}. Within this framework, one defines a commuting family of transfer matrices dependent on an affine spectral parameter. For spin chains, the Hamiltonian is recovered from the logarithmic derivative of the transfer matrix evaluated at a specific value of the spectral parameter-typically zero. The spectral problem then reduces to diagonalizing these transfer matrices. For quantum systems with internal degrees of freedom, this diagonalization becomes an interacting many-body problem within the internal space, ultimately requiring a nested Bethe ansatz. In addition to the particle momenta, one introduces spin rapidities which describe the collective spin
or isospin degrees of freedom. This nested structure is central to the static $SU(2)$ Gross-Neveu
and $U(1)$-symmetric Thirring models.  In both cases the exact eigenstates
of the constant-coupling Hamiltonian are obtained by a nested Bethe
ansatz:  the particles are ordered in the configuration space which
have definite momenta, and the interactions between the particles
gives rise to S-matrices which act in the internal spin spaces of the particles.  In the chiral-invariant
$SU(2)$ Gross-Neveu model the two-particle scattering matrix is associated with the rational XXX $R$-matrix, while
in the anisotropic $U(1)$-symmetric Thirring model it is replaced by
the trigonometric or hyperbolic XXZ $R$- matrices. These S-matrices satisfy the Yang-Baxter equations, which guarantees consistency
of the wavefunction. Applying periodic boundary conditions results in the emergence of a transfer matrix, 
which is then diagonalized resulting in coupled Bethe equations involving the momenta and the spin rapidities of the particles
\cite{AndreiLowenstein79,JaparidzeNersesyanWiegmann1984,Andrei80,Wiegmann_1981}.  

Introducing explicit time dependence into such an interacting quantum
field theory changes the problem fundamentally.  The eigenstates of the
instantaneous Hamiltonian $H(t)$ do not in general solve the
time-dependent Schrodinger equation. One must therefore seek an exact solution to the time-dependent Schrodinger equation. 
Naturally, integrability imposes restrictions on the allowed temporal dependence of the Hamiltonian, which becomes part of the integrability problem.

Important classes of exactly solvable time-dependent Hamiltonians had
previously been constructed using Gaudin Hamiltonians that are associated with classical Yang-Baxter algebra
\cite{TCSun,Patra_2015,Sinitsyn2018,Yuzbashyan2018}.  These include multilevel
Landau-Zener systems and collective or reduced models related to
Gaudin magnets and BCS dynamics. More recently, this line of work has revealed an intriguing connection
between exactly solvable time-dependent quantum mechanics and
Painlev\'e integrability.  Sinitsyn studied an integrable multivariable
generalization of the Painlev\'e-II equation and used its Lax
representation, together with the exact solution of the
Demkov-Osherov time-dependent quantum model, to derive connection
formulas relating asymptotic solutions at different infinities
\cite{SinitsynPainleve2026,SinitsynPainleveN2026}. These results provide a direct link
between solvable time-dependent quantum-mechanical models and
Painlev\'e-type classical integrable systems.
Such constructions provide important examples of exact time-dependent quantum evolution, but they concern a
different class of systems from the interacting quantum field theories
solved by the conventional Bethe ansatz which involves factorized many-particle 
scattering and quantum Yang-Baxter equations. Although the underlying construction remains distinct from the Bethe ansatz for interacting quantum field theories, they are conceptually
relevant to the broader mathematical structures discussed later in this
paper.

The generalized Bethe ansatz was introduced to address precisely the
problem of solving quantum field theories with time-dependent coupling strengths \cite{PasnooriKondo2025,PasnooriQKZ2026}.  It extends the
Bethe-ansatz construction to interacting many-body systems and quantum
field theories with time-dependent interaction strengths.  As in the
ordinary coordinate Bethe ansatz, the wavefunction is decomposed into
amplitudes associated with different orderings of particles in
configuration space, and the amplitudes are related by two-particle
scattering matrices.  The essential difference is that the amplitudes
and scattering matrices are now spacetime dependent.  Requiring the
many-body wavefunction to be independent of the sequence of particle
exchanges gives the Yang-Baxter consistency conditions and constrains
the allowed time dependence of the couplings.  Periodic boundary
conditions then produce matrix difference equations whose
solutions determine the exact wavefunction satisfying the
time-dependent Schrodinger equation \cite{PasnooriQKZ2026,pasnoorikondo2}. This method as recently been extended to the case of time dependent boundary conditions \cite{pasnooriBoundary2026} and and also to the case of non-Hermitian models \cite{pasnooriNHKtime}. For models with an internal spin or isospin degree of freedom, the
generalized Bethe ansatz is nested.  In the time-dependent
$SU(2)$ Gross-Neveu model, periodic transport gives rise to quantum Knizhnik-Zamolodchikov (qKZ) equations
associated with the rational XXX $R$-matrix
\cite{PasnooriQKZ2026}.  In the anisotropic $U(1)$-symmetric Thirring
model the corresponding construction gives trigonometric qKZ equations
associated with the XXZ $R$-matrix.  The amplitudes are vector valued in
the internal spin space, and their off-shell Bethe-ansatz solutions
involve auxiliary rapidities and Jackson-type sums
\cite{FrenkelReshetikhin1992,Reshetikhin1992,
BabujianKarowskiZapletal1997}.  Thus, in these chiral theories, the
nontrivial finite-difference structure enters through many-body
transport in the internal spin sector.

A remarkable property of these time-dependent integrable systems is
their relation to renormalization-group flow.  Hoare, Levine and
Tseytlin showed that for a class of two-dimensional integrable field
theories, promoting couplings to spacetime-dependent functions while
preserving a Lax representation constrains the allowed dependence
according to the RG flow of the corresponding static theory
\cite{HoareLevineTseytlin2020}.  The generalized Bethe ansatz provides
a quantum realization of the same phenomenon.  Here the constraint
arises from consistency of the exact time-dependent many-body
wavefunction, and in the universal regime the resulting equations for
the couplings coincide with the RG equations when physical time is
identified, up to normalization, with the logarithmic RG scale
\cite{PasnooriRG2026,PasnooriTransmutation}.  In theories with cyclic RG trajectories this
correspondence becomes especially striking: the RG limit cycle is
realized as a periodic trajectory of the interaction strengths in
physical time \cite{PasnooriBKT2026}.

There is also a complementary gauge-theoretic perspective on these
structures.  Costello, Witten and Yamazaki showed that Yang-Baxter
structures, $R$-matrices and associated quantum-integrable systems can
be obtained from four-dimensional Chern-Simons theory
\cite{Costello1,Costello2,costello3}; see also
Ref.~\cite{Yamazaki2025Overview} for a recent overview of the relation
between gauge theory and integrability.  More recently, Komatsu,
Sakamoto, Wallberg and Yamazaki generalized this framework to
spacetime-dependent data and constructed time-dependent classically
integrable field theories whose allowed coupling dependence reproduces
the corresponding one-loop RG evolution
\cite{KomatsuSakamotoWallbergYamazaki2026}.  These results provide a
higher-dimensional gauge-theoretic realization of the
integrability-RG correspondence and make it natural to ask how the
finite-difference transport and exact quantum states arising in the
generalized Bethe ansatz fit into the same broader framework.

The family of current-current deformations of the
$SU(2)_1$ Wess-Zumino-Witten model provides a particularly useful
setting in which to compare these approaches.  Non-Abelian bosonization
separates a doublet of massless Dirac fermions into a free $U(1)$ charge
sector and an interacting $SU(2)_1$ spin sector
\cite{Witten1984,JaparidzeNersesyanWiegmann1984}.  An anisotropic
current-current deformation of the latter is equivalently described by
the interacting spin sector of the $U(1)$-symmetric Thirring model.
Abelian bosonization gives the sine-Gordon description, while
refermionization gives the massive Thirring model in the corresponding
massive regime \cite{Coleman1975,Mandelstam1975}.  The isotropic
current-current deformation gives the chiral invariant
$SU(2)$ Gross-Neveu model.  The same interacting degrees of freedom
may therefore be accessed through rather different integrability
constructions: qKZ equations in the spinful fermionic
description, a spacetime-dependent Lax connection in the sine-Gordon
description, and, as we show here, a direct coordinate generalized
Bethe ansatz in the massive-Thirring representation.

The massive Thirring model considered in the present work differs in an
important way from the chiral models previously treated by the
generalized Bethe ansatz.  In the $SU(2)$ Gross-Neveu and anisotropic
$U(1)$ Thirring models, the elementary particles are massless chiral
fermions carrying an internal spin degree of freedom.  Their
one-particle equations are simple chiral transport equations, and the
nontrivial structure lies in the many-body spin amplitudes.  Periodic
transport acts on this internal space and gives matrix qKZ equations,
whose solution requires the off-shell Bethe ansatz \cite{BabujianKarowskiZapletal1997}. The massive theory considered here has a fundamentally different
structure because the explicitly time-dependent bare mass couples the
right- and left-moving components already in the one-particle sector.
Rather than assuming a particular temporal protocol, we require the
existence of a consistent rapidity-labelled generalized coordinate
Bethe wavefunction.  These consistency conditions select
\begin{equation}
  m_b(t)=m_*e^{\kappa t},
\label{eq:intro_mass_protocol}
\end{equation}
so that the exponential mass protocol emerges directly from the exact
coordinate construction.  As we discuss below, this time dependence
has precisely the form of the RG evolution of the mass parameter upon
identifying the physical time $t$ with the logarithm of the cutoff.
The present model therefore realizes the integrability-RG
correspondence in a different manner from the previously studied
time-dependent chiral theories: here the correspondence is already
encoded in the compatibility of the one-particle massive dynamics.

Since the massive-Thirring fermions carry no independent isospin
degree of freedom, the amplitudes entering the coordinate wavefunction
are scalar functions.  Periodic boundary conditions act directly on
the physical rapidities and introduce a new spectral structure: a
single rapidity sector is not closed under spatial winding.  Instead,
winding a particle around the circle shifts its rapidity by a finite
amount, so periodicity of the complete state requires a discrete
rapidity lattice.  The physical wavefunction must therefore include a
sum over all rapidity sectors generated by these finite shifts.  In
this sense, the present construction differs from the nested
generalized Bethe ansatz, where finite-difference transport acts on
vector-valued amplitudes and auxiliary rapidities in an internal spin
space.

The interacting many-body structure enters once the different
configuration-space orderings are assembled.  The quartic fermionic
interaction produces a scalar two-particle S-matrix relating
the amplitudes associated with neighboring particle orderings.
Combining these exchange relations with periodic winding leads to
scalar analytic difference equations of Ruijsenaars type, that relate amplitudes at
neighboring points of the physical rapidity lattice.  Their ratios are
fixed by the massive-Thirring S-matrix, and the corresponding
solutions can be expressed in terms of $q$-gamma functions.  The
finite-difference step, and hence the associated $q$ parameter, is
determined by the drive rate and the system size.  The complete
$N$-particle solution is obtained by combining these
amplitudes corresponding to different orderings of the particles with the full rapidity-lattice sum. In this sense, the present
theory provides a generalized coordinate Bethe-ansatz realization in
which the finite-difference transport acts directly on the physical
rapidities rather than on auxiliary rapidities in an internal nested
space.

The construction also has the correct static limit.  When the explicit
time dependence is removed, the finite rapidity translation collapses
to the familiar phase accumulated by a massive particle moving around
the system, and the finite-difference periodicity conditions
reduce to the standard massive-Thirring Bethe equations.  Thus the
rapiditiy-lattice formulation provides a genuine time-dependent
generalization of the conventional coordinate Bethe ansatz.

There exists another exact representation, which is obtained by resolving the
one-particle Dirac equation into eigenstates of spatial momentum.  For
the exponential mass protocol the resulting time-dependent equations
reduce to Whittaker equations.  The rapidity and momentum
representations therefore expose two different special-function
structures of the same exact solution: $q$-gamma functions associated
with finite spectral transport and ordinary Whittaker functions
associated with momentum-resolved time evolution.  We also discuss how
the analytic difference equations appearing here are related to
mathematical structures surrounding qKZ equations, quantum affine
algebras, Macdonald-Ruijsenaars systems, $q$-Whittaker functions,
$q$-Painlev\'e tau functions and gauge theory.  Throughout that
discussion we distinguish established correspondences from structural
similarities for which no explicit map is presently known.

Broadly speaking, the current work has two purposes: The first purpose is to develop a generalized
\emph{coordinate} Bethe-ansatz construction for a time-dependent
relativistic quantum field theory whose integrable structure is
different from that of the previously studied chiral models using nested generalized Bethe ansatz. The second purpose is to establish the time-dependent integrability-RG correspondence
in a different setting compared to previously studied
chiral theories using nested generalized Bethe ansatz and the Lax formalism of Hoare, Levine and Tseytlin. 

The paper is organized as follows.  In
Sec.~\ref{sec:model} we introduce the massive Thirring model and review
its relation to the $U(1)$-symmetric Thirring model, the anisotropic
current-current deformation of the $SU(2)_1$ WZW model and the
sine-Gordon model.  In Sec.~\ref{sec:construction} we solve the
one-particle problem, derive the allowed exponential mass protocol and
show how periodic boundary conditions generate the physical rapidity
lattice.  We then construct the two-particle coordinate wavefunction
and derive the scalar S-matrix.  In Sec.~\ref{sec:solution}
we generalize the construction to arbitrary particle number and solve
the resulting analytic difference equations.  We subsequently discuss
the static Bethe limit and the momentum-space Whittaker
representation.  The relation to qKZ equations, Ruijsenaars-type
difference equations, $q$-Whittaker and $q$-Painlev\'e structures, and
gauge-theoretic constructions is discussed in
Sec.~\ref{sec:mathcontext}.  We conclude in
Sec.~\ref{sec:conclusion} with a summary and directions for future
work.

\section{Massive Thirring model and the related field theories with internal degrees of freedom}
\label{sec:model}

We consider a Dirac fermion on a circle of circumference $L$ and set the velocity and $\hbar$ to one. The Hamiltonian is 
\begin{align}
 H(t)=\int_0^L\dd x\,\Big[&-\ii\psi_R^\dagger\partial_x\psi_R +\ii\psi_L^\dagger\partial_x\psi_L +\ii\mb(t)(\psi_R^\dagger\psi_L-\psi_L^\dagger\psi_R) \nonumber\\&+2g_0\psi_R^\dagger\psi_L^\dagger\psi_L\psi_R\Big].
\label{eq:H}
\end{align}
The fields satisfy canonical equal-time anticommutation relations and periodic boundary conditions,
\begin{equation}
 \{\psi_a(x),\psi_b^\dagger(y)\}=\delta_{ab}\delta_L(x-y),
 \qquad \psi_a(x+L)=\psi_a(x),\qquad a,b=R,L.
\label{eq:fields}
\end{equation}
The bare mass $\mb(t)$ is real and spatially homogeneous, whereas $g_0$ is a real constant. We follow the original coupling and conventions of Bergknoff and Thacker \cite{BergknoffThacker1979,Thacker1981}. We use the following spectral convention
\begin{equation}
 \cot\mu=-\frac{g_0}{2},\qquad \mu=\frac\pi2+\arctan\frac{g_0}{2},\qquad 0<\mu<\pi.
\label{eq:mu}
\end{equation}
For $\mu=\pi/2$, the quartic fermion interaction strength vanishes and the model is that of non interacting fermions. This is referred to as the Luther-Emery line. Particle-antiparticle repulsion corresponds in this convention to $g_0<0$, or $\mu<\pi/2$, and attraction to $g_0>0$, or $\mu>\pi/2$. These labels refer to the excitations relative to the physical vacuum in the case of constant strengths. They should not be inferred solely from the sign of an interaction between bare fermions. We will distinguish $\mu$ from the crossing parameter of the spinful $U(1)$ Thirring model whenever both appear.

Let $\epsilon(r)$ be the sign function. The distributional prescription is
\begin{equation}
 \partial_r\epsilon(r)=2\delta(r),\qquad \delta(r)\epsilon(r)=0.
\label{eq:prescription}
\end{equation}
The second identity is the symmetric prescription of Ref.~\cite{BergknoffThacker1979}. 
Just as in the chiral models, the number of particles is a conserved quantity, where
\begin{equation}
 N=\int_0^L\dd x\,(\psi_R^\dagger\psi_R+\psi_L^\dagger\psi_L).
\label{eq:number}
\end{equation}
But unlike the chiral models, the mass term does not preserve the two chiral particle numbers separately. We thus construct the wavefunction in a sector of fixed number of particles $N$ above the empty reference state $\ket0$, defined by $\psi_R(x)\ket0=\psi_L(x)\ket0=0$. As in the static coordinate solution, this reference state is not the physical vacuum. In the static theory the physical vacuum is obtained only after filling the negative-energy branch and constructing the renormalized excitation spectrum \cite{BergknoffThacker1979,Thacker1981,PasnooriMT2025,PasnooriSGPT}.

\begin{center}
\begin{figure}
\includegraphics[width=0.9\columnwidth]{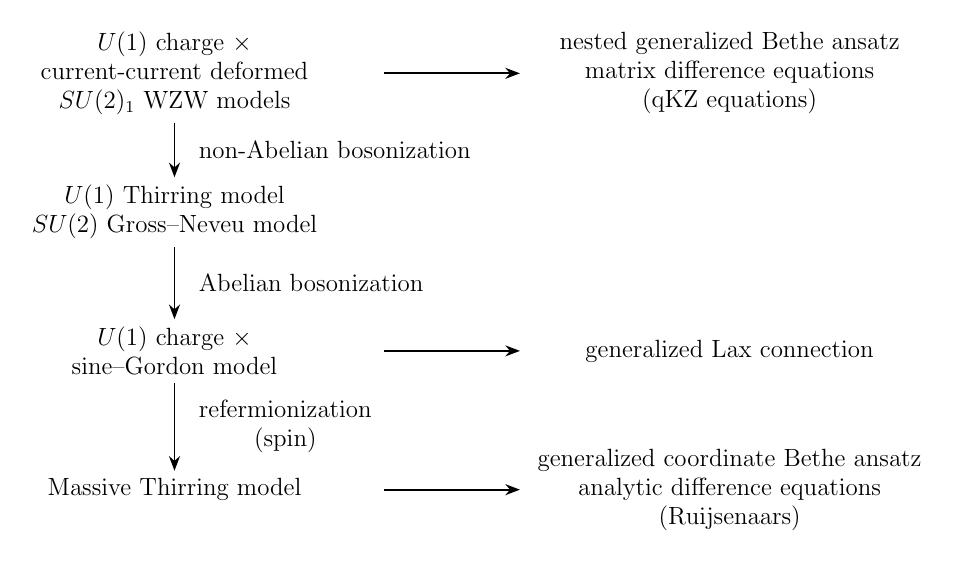}
\caption{
Schematic relation between the field-theory descriptions and the
corresponding integrability constructions discussed in this work.
Non-Abelian bosonization relates the $U(1)$-symmetric Thirring model
to a decoupled $U(1)$ charge sector and an anisotropic current-current
deformation of the $SU(2)_1$ WZW model; the isotropic limit gives the
$SU(2)$ Gross-Neveu model.  Abelian bosonization of the interacting
spin sector gives the sine-Gordon model, while refermionization of
this sector gives the massive Thirring model.  These equivalent
descriptions organize time-dependent integrability in different ways.
For the $U(1)$ Thirring and $SU(2)$ Gross-Neveu models, the
generalized Bethe ansatz is nested and periodic transport leads to
matrix qKZ equations.  In the sine-Gordon description, integrability
is formulated through a generalized spacetime-dependent Lax
connection.  In the massive Thirring representation developed here,
the generalized coordinate Bethe ansatz acts directly on the physical
rapidities and leads to scalar analytic difference equations of
Ruijsenaars type.  The figure emphasizes that the same interacting
sector admits distinct but related integrable descriptions.}
\label{fig:picture1}
\end{figure}
\end{center}

The models discussed in this work and in our earlier studies are related
by standard bosonization correspondences. Since the same interacting spin sector appears in several
different representations, it is useful to summarize
these relations before constructing the time-dependent massive-Thirring
wavefunction. We begin with the $U(1)$-symmetric Thirring model containing a doublet
of right- and left-moving fermions,
$\Psi_{Ra}$ and $\Psi_{La}$, with $a=\uparrow,\downarrow$.  Non-Abelian
bosonization separates the charge and spin degrees of freedom.  The
charge sector is a free compact $U(1)$ boson, whereas the spin sector is
described by the level-one $SU(2)_1$ Wess-Zumino-Witten model
\cite{Witten1984}.  The anisotropic four-fermion interaction of the
fermionic theory becomes an anisotropic current-current deformation of
the WZW model,
\begin{align}
 H_{\rm s}(t)=H_{SU(2)_1} -8\int_0^L\dd x\,\big[&g_\perp(t)(J_R^xJ_L^x+J_R^yJ_L^y)+g_\parallel(t)J_R^zJ_L^z \big].
\label{eq:WZWspin}
\end{align}
The charge sector is a spectator in the discussion below. The isotropic limit, $g_\parallel=g_\perp$, restores the full $SU(2)$ symmetry of the interacting spin sector which corresponds to the chiral-invariant $SU(2)$ Gross-Neveu model. %Away from this limit the interaction is anisotropic and the associated S-matrices correspond to the trigonometric and hyperbolic XXZ R-matrices rather thanthe rational XXX R-matrices corresponding to the $SU(2)$ case \cite{JaparidzeNersesyanWiegmann1984,PasnooriQKZ2026,PasnooriBKT2026}.
The same anisotropic spin sector also admits an Abelian bosonized
description.  In the massive regime it is represented by the
sine-Gordon model,
\begin{equation}
 H_{\mathrm{SG}}(t)=\int_0^L\dd x\left[ \frac12\Pi^2+\frac12(\partial_x\Phi)^2+\lambda_{\mathrm{SG}}(t): \cos(\beta_{\mathrm{SG}}\Phi): \right].
\label{eq:SG}
\end{equation}
Here $\Phi(x)$ is a compact bosonic field and $\Pi(x)$ is its canonically conjugate momentum,
\begin{equation}
 [\Phi(x),\Pi(y)]=i\delta(x-y),
\end{equation}
with all other equal-time commutators vanishing.  The field $\Phi$ is identified modulo its compactification period.
One may introduce the dual field $\Theta$ through
$\Pi\propto\partial_x\Theta$, with the precise normalization fixed by
the bosonization convention. Depending on the value of $\beta_{\mathrm{SG}}$, the sine-Gordon cosine term then gaps the
compact bosonic sector by pinning $\Phi$ to one of its minima. The coupling strength corresponding to the longitudinal current-current deformation in (\ref{eq:WZWspin}) is related to the parameter
$\beta_{\mathrm{SG}}$ in the SG model (\ref{eq:SG}), while the transverse interaction corresponds to the cosine parameter.  The word ``deformation'' is used here in
the finite-coupling sense; no weak-coupling expansion about the $SU(2)_1$ point is implied. The sine-Gordon model is in turn related by the Coleman-Mandelstam
fermion-boson correspondence to the massive Thirring model
\cite{Coleman1975,Mandelstam1975}.  In a fixed operator normalization,
the fermion mass operator and the sine-Gordon vertex are related
schematically by
\begin{equation}
 \ii(\psi_R^\dagger\psi_L-\psi_L^\dagger\psi_R) \quad\longleftrightarrow\quad \mathcal C_{\rm UV}:\cos(\beta_{\mathrm{SG}}\Phi): ,\qquad\lambda_{\mathrm{SG}}(t)=\mathcal C_{\rm UV}\,\mb(t).
\label{eq:massbosonization}
\end{equation}
The coefficient $\mathcal C_{\rm UV}$ depends on the ultraviolet
normalization and regularization and should not be identified directly
with a physical soliton mass. The relations relevant for the present discussion may therefore be
summarized as shown in the Fig.~(\ref{fig:picture1}).
These equivalent descriptions have been useful for studying
time-dependent integrability from complementary directions.  In our
earlier work on the time-dependent $SU(2)$ Gross-Neveu model, the
generalized Bethe ansatz leads to rational qKZ equations associated
with the XXX $R$-matrix.  In the anisotropic $U(1)$-symmetric Thirring
model, or equivalently the anisotropic current-current deformation of
the $SU(2)_1$ WZW spin sector, the same construction leads instead to
trigonometric and hyperbolic qKZ equations associated with the XXZ $R$-matrix
\cite{PasnooriQKZ2026,PasnooriBKT2026}.  In both cases the qKZ
equations determine the spacetime-dependent amplitudes entering the
exact many-body wavefunction, and their consistency constrains the
allowed time dependence of the couplings.

A complementary approach was developed by Hoare, Levine and Tseytlin
for the classical sine-Gordon and related sigma-model descriptions
\cite{HoareLevineTseytlin2020}.  There the starting point is a
space-time dependent classical Lax connection.  Requiring zero
curvature constrains the local couplings and, for the spatially homogeneous
sine-Gordon model, selects the same type of exponential
time dependence (\ref{eq:intro_mass_protocol}) associated with the RG flow of the static model. Thus, their construction
establishes the integrability-RG connection at the level of the
classical Lax formalism.

The present massive-Thirring problem provides a different realization of the integrability-RG flow connection. In contrast to the chiral models mentioned above which requires a nested generalized Bethe ansatz, here since the fermions are non-chiral and do not have an internal spin space, we work  directly with the coordinate wavefunction of the massive fermions. Compatibility of a rapidity-labelled one-particle family selects the allowed time dependence of the mass; the quartic fermionic interaction results in a scalar S-matrix; and periodic boundary conditions produce analytic difference equations for the physical rapidities themselves.  The resulting construction may therefore be viewed as a direct coordinate version of the generalized Bethe ansatz adapted to the massive Thirring model.

\section{Generalized coordinate Bethe wavefunction}
\label{sec:construction}

Having introduced the massive Thirring model and its relation to the
spinful chiral models and to the sine-Gordon model, we now turn
to the exact solution of the time-dependent Schrodinger equation by
means of a generalized coordinate Bethe ansatz.  We begin with the
one-particle sector, where the coupled right- and left-moving components
form a rapidity-labelled family whose compatibility determines the
admissible time dependence of the bare mass.  We then construct the
two-particle wavefunction by resolving configuration space into the two
possible orderings of the particles with respect to each other. The quartic fermion interaction fixes the relation between the amplitudes corresponding to each ordering through a scalar S-matrix.  Finally, the
$N$-particle state is assembled from these one- and two-particle
building blocks: amplitudes associated with different
orderings of the particles are related by products of scalar S-matrices, while
periodic boundary conditions require a sum over the physical rapidity
lattice and lead to first-order analytic difference equations of
Ruijsenaars type which impose constraints on the corresponding amplitudes. Solving these
analytic difference equations determines the amplitudes on the rapidity
lattice and thereby provides the complete time-dependent many-body
wavefunction.

\subsection{One-particle sector}
\label{sec:one}
We begin with the one-particle sector. Unlike in the chiral models previously treated by the generalized Bethe ansatz, due to the presence of the time-dependent mass, the solution is already nontrivial at this level. We first solve the time-dependent Dirac equation and ask under what conditions it admits a rapidity-labelled family of spinors compatible with the coordinate Bethe construction.  This requirement will constrain the allowed time dependence of the bare mass and select the exponential protocol.  We then impose periodic boundary conditions on the one-particle state.  Rather than quantizing a fixed rapidity, spatial winding will be shown to shift the rapidity by a finite amount, thereby requiring the introduction of the physical rapidity lattice that forms the basis of the subsequent many-body construction.

\subsubsection{The one-particle equation and rapidity ansatz}

The one-particle state is written directly in the spectral
normalization that will be used throughout the finite-step periodic
construction,
\begin{equation}
 \ket{1;t}=\int_0^L\dd x\, \left[ \widehat f^R(x,t)\psi_R^\dagger(x)+\widehat f^L(x,t)\psi_L^\dagger(x)\right]\ket0 .
\end{equation}
The quartic interaction annihilates a one-particle state.  Acting with
the quadratic part of the Hamiltonian and comparing the coefficients
of the two creation operators gives
\begin{equation}
 \ii\partial_t\begin{pmatrix} \widehat f^R\\ \widehat f^L\end{pmatrix} =
 \begin{pmatrix}
  -\ii\partial_x&\ii\mb\\-\ii\mb&\ii\partial_x
 \end{pmatrix}
 \begin{pmatrix}
  \widehat f^R\\ \widehat f^L
 \end{pmatrix}.
\label{eq:Diracmatrix}
\end{equation}
Equivalently,
\begin{equation}
 (\partial_x+\partial_t)\widehat f^R=\mb(t)\widehat f^L,\qquad(\partial_x-\partial_t)\widehat f^L=\mb(t)\widehat f^R.
\label{eq:Dirac}
\end{equation}

Introduce the null coordinates
\begin{equation}
 z=x-t,\qquad \bar z=x+t,\qquad \partial_x+\partial_t=2\partial_{\bar z},\qquad\partial_x-\partial_t=2\partial_z .
\label{eq:null}
\end{equation}
Unlike in the massless chiral theories, the two components of a
massive one-particle solution are coupled and therefore generally
depend on both null coordinates.  We seek a continuous
rapidity-labelled family of the form
\begin{equation}
 \widehat f_\beta^R =e^{\beta/2}F(z)\widehat G_\beta(z,\bar z),\qquad \widehat f_\beta^L =-\ii e^{-\beta/2}\bar F(\bar z) \widehat G_\beta(z,\bar z),
\label{eq:rapidityansatz}
\end{equation}
where $F$ and $\bar F$ are independent of $\beta$.  The bar on
$\bar F$ labels its dependence on $\bar z$ and does not denote complex
conjugation.  The factors $e^{\pm\beta/2}$ are the familiar relative
spinor weights of the static massive Dirac problem.  The question is
whether they can be retained as rapidity labels when the mass becomes
explicitly time dependent. Since

\begin{equation}
 \partial_{\bar z}F=0, \qquad \partial_z\bar F=0,
\end{equation}
Eq.~\eqref{eq:Dirac} reduces to
\begin{equation}
 2\partial_{\bar z}\log \widehat G_\beta = -\ii e^{-\beta}\mb\frac{\bar F}{F}, \qquad 2\partial_z\log \widehat G_\beta = \ii e^\beta\mb\frac{F}{\bar F}.
\label{eq:Geq}
\end{equation}
For a continuous rapidity family, equality of the mixed derivatives
requires
\begin{equation}
 -e^{-\beta} \partial_z\!\left(\mb\frac{\bar F}{F}\right)= e^\beta \partial_{\bar z}\!\left(\mb\frac{F}{\bar F}\right).
\end{equation}
Since this relation must hold for all $\beta$, the coefficients of
$e^{-\beta}$ and $e^\beta$ must vanish separately.  Hence there exist
functions $a_+(\bar z)$ and $a_-(z)$ such that
\begin{equation}
 \mb\frac{\bar F}{F}=a_+(\bar z), \qquad \mb\frac{F}{\bar F}=a_-(z),\qquad \mb^2\!\left(\frac{\bar z-z}{2}\right)=a_+(\bar z)a_-(z).
\label{eq:massfactorization}
\end{equation}

To see the implication explicitly, take the logarithm of the absolute
value of the last equality:
\begin{equation}
 2\log\left| \mb\!\left(\frac{\bar z-z}{2}\right) \right|=\log|a_+(\bar z)|+ \log|a_-(z)|.
\end{equation}
The mixed derivative of the right-hand side vanishes.  Since
\begin{equation}
 \partial_z t=-\frac12, \qquad \partial_{\bar z}t=\frac12,
\end{equation}
the same derivative of the left-hand side gives
\begin{equation}
 \partial_z\partial_{\bar z} \left[2\log|\mb(t)|\right] = -\frac12\frac{\dd^2}{\dd t^2}\log|\mb(t)| =0.
\label{eq:mixedmass}
\end{equation}
Thus the logarithm of a smooth nonzero mass is affine in time, with
its fixed sign absorbed into the integration constant.

Within the smooth nonzero rapidity-factorized family
\eqref{eq:rapidityansatz}, compatibility of the one-particle equations
for all rapidities therefore requires, and is satisfied by,
\begin{equation}
 \mb(t)=m_*e^{\kappa t}.
\label{eq:protocol}
\end{equation}
The constants $m_*$ and $\kappa$ are independent of the rapidity
label, and a shift of the time origin may be absorbed into $m_*$.  This
result is a restriction on the one-particle building blocks required
by the generalized coordinate Bethe ansatz; it is not a no-go theorem
for arbitrary time-dependent Dirac equations.  In particular, for
$g_0=0$ the theory is free for an arbitrary sufficiently regular
$\mb(t)$.  The exponential protocol is selected by requiring the
uniform rapidity-labelled family used in the interacting many-body
construction.  Since the one-particle equations contain no $g_0$,
this analysis does not determine a relation $\kappa(g_0)$.

\subsubsection{Explicit form of the one-particle solution}

A choice of the null factors that solves Eq.~\eqref{eq:massfactorization} is
\begin{equation}
 F(z)=e^{-\kappa z/2},\qquad \bar F(\bar z)=e^{\kappa\bar z/2}.
\end{equation}
We fix the rapidity-dependent integration constant so that spatial winding will act as a pure translation of the rapidity label. Integrating Eq.~\eqref{eq:Geq} then gives
\begin{equation}
 \widehat G_\beta= \exp\!\left[-\frac{\ii m_*}{2\kappa}\left\{e^{-\beta}e^{\kappa\bar z} +e^\beta e^{-\kappa z}\right\}\right].
\label{eq:Gexplicit}
\end{equation}
Consequently,
\begin{equation}
 \begin{pmatrix}\widehat f_\beta^R\\ \widehat f_\beta^L\end{pmatrix}=e^{\kappa t/2}
 \begin{pmatrix}e^{(\beta-\kappa x)/2}\\-\ii e^{-(\beta-\kappa x)/2}\end{pmatrix}
 \exp\!\left[-\frac{\ii}{\kappa} \mb(t)\cosh(\beta-\kappa x)\right].
\label{eq:fhat}
\end{equation}
Direct substitution verifies both equations in Eq.~\eqref{eq:Dirac}.  Equation~\eqref{eq:fhat} is the elementary one-particle block used throughout the exact time-dependent construction.  Its normalization is chosen to make periodic transport transparent, rather than to make the pointwise $\kappa\to0$ limit manifest.  The plane-wave normalization appropriate to the conventional static Bethe limit will be introduced only in Sec.~\ref{sec:staticlimit}.

\subsubsection{Spatial winding and the rapidity lattice}

A translation $x\mapsto x+L$ shifts the combination $\beta-\kappa x$ by
\begin{equation}
 p=\kappa L.
\label{eq:step}
\end{equation}
Equation~\eqref{eq:fhat} then gives directly
\begin{equation}
 \widehat f_\beta(x+L,t)=\widehat f_{\beta-p}(x,t).
\label{eq:hatwinding}
\end{equation}
Thus a spatial winding does not return a fixed-rapidity block to itself; it translates the physical rapidity label by one finite step.  The naive fixed-rapidity condition $ \widehat f_\beta(x+L,t)=\widehat f_\beta(x,t)$
cannot therefore be imposed for generic $\beta$.  Periodicity of the complete one-particle state instead requires the full rapidity lattice
\begin{equation}
 \beta=\widetilde\beta+pn, \qquad n\in\mathbb Z.
\label{eq:oneparticlelattice}
\end{equation}
The rapidity lattice is thus not introduced in order to solve a later difference equation; it is already forced by the one-particle dynamics together with periodic boundary conditions.

This is fundamentally different from the massless chiral models studied in the earlier generalized Bethe-ansatz constructions.  There the one-particle equations are chiral transport equations, so the functions of the null coordinates remain arbitrary until many-body transport and boundary conditions are imposed.  Here the time-dependent mass couples the two chiral components and fixes the nontrivial $x$-, $t$- and rapidity-dependence of the one-particle spinor.  The lattice of physical rapidities is therefore present before interactions are considered.  In the many-body problem due to the interaction term, the scattering between particles does something additional: it determines the relative weights carried by neighboring points of these already-existing rapidity lattices.

\subsection{The exponential protocol, RG evolution and the sine-Gordon}
\label{sec:HLT}

The condition just obtained has a simple RG interpretation. On the line of Luttinger liquid fixed points, the normal-ordered sine-Gordon vertex has scaling dimension $\Delta=\beta_{\mathrm{SG}}^2/(4\pi)$ in the convention of Eq.~\eqref{eq:SG} \cite{Coleman1975}. The dimensionless coupling $h$ has the following RG equation
\begin{equation}
 \frac{\dd h}{\dd\ell}=y_m h, \qquad h(\ell)=h(\ell_*)e^{y_m(\ell-\ell_*)}.
\label{eq:massRG}
\end{equation}
Here $y_m=2-\Delta$ and $\ell$ increases toward long distances. The operator exponent is fixed by the constant compactification parameter. At a fixed ultraviolet normalization, Eq.~\eqref{eq:massbosonization} makes $h(t)$ proportional to $\mb(t)$. Thus the GBA-selected time dependence has the multiplicative form of the mass direction of the RG trajectory. For $y_m\ne0$, the identification $\ell-\ell_*=(\kappa/y_m)(t-t_*)$ matches these parameterizations. It is important that Eq.~\eqref{eq:massRG} has not been used to construct the spinor. The exponential followed from compatibility of the one-particle family. Moreover, the local equation does not determine the normalization of RG time or a coupling-dependent value of $\kappa$. 

A closely related condition was found in Appendix E of Hoare, Levine and Tseytlin \cite{HoareLevineTseytlin2020}. To avoid confusing their variables with ours, write their Lagrangian as
\begin{equation}
 \mathcal L_{\mathrm H}=\frac1{g_{\mathrm H}^2} \left[\frac12\partial_+X\partial_-X+m_{\mathrm{SG}}^2\cos X\right].
\label{eq:HLTL}
\end{equation}
Their Lax matching keeps $g_{\mathrm H}$ constant and requires
\begin{equation}
 m_{\mathrm{SG}}^2(\tau,\sigma) =v_+(\xi^+)v_-(\xi^-),\qquad \xi^\pm=\frac{\tau\pm\sigma}{2}.
\label{eq:HLTfactorization}
\end{equation}
They use a conformal transformation to select the homogeneous representative
\begin{equation}
 m_{\mathrm{SG}}^2(\tau)=m_{\mathrm{SG},0}^2e^{\beta_{\mathrm H}\tau}, \qquad \beta_{\mathrm H}=-2+g_{\mathrm H}^2,
\label{eq:HLTbeta}
\end{equation}
identified there with one-loop RG time in their normalization. Their Eq.~(E.5) gives the corresponding flat Lax connection and shows that a spatial translation can be absorbed into a rescaling of the spectral parameter \cite{HoareLevineTseytlin2020}. The separated dependence on the two null coordinates in Eq.~\eqref{eq:HLTfactorization} is the bosonic analogue of the factorization condition derived in Eq.~\eqref{eq:massfactorization}. In the present formulation, however, this condition emerges directly from the consistency of the quantum wavefunction. Through bosonization, the coefficient of the sine-Gordon potential is related, up to a fixed operator normalization, to the fermionic mass parameter. Equations~\eqref{eq:protocol} and \eqref{eq:HLTbeta} therefore exhibit the same qualitative exponential dependence on time. A quantitative comparison of the corresponding rates requires fixing both the normalization of the bosonic field and the relation between the time coordinates $t$ and $\tau$. We thus identify the same underlying integrability-RG mechanism in the two descriptions, without equating $\kappa$ and $\beta_{\mathrm H}$ purely by notation.

The subsequent two-particle calculation is entirely quantum and keeps the contact interaction at finite $g_0$. It will show that the same protocol supports a scalar many-body construction. This provides information not contained in the local Lax-factorization condition: the explicit state, its scattering amplitudes, and the spectral transport associated with periodic boundary conditions.

\subsection{Two-particle sector}
\label{sec:two}

We now consider the two-particle sector. As in the standard coordinate
Bethe ansatz, the interaction acts when the two particles cross and
therefore relates the amplitudes associated with the two possible
orderings in configuration space. It is useful to introduce these amplitudes explicitly from the outset. Let
\begin{equation}
 \bbet=(\beta_1,\beta_2), \qquad \theta=\beta_1-\beta_2 ,
\end{equation}
and denote the two amplitudes corresponding to different orderings of the particles by
$A_{12}(\beta_1,\beta_2)$ and $A_{21}(\beta_1,\beta_2)$.  From this point onward these amplitudes are defined relative to the spectrally normalized one-particle blocks $\widehat f$.  This convention removes the purely one-particle winding phase from the finite-step transport equations.
The two-particle state is written as
\begin{equation}
 \ket{\beta_1,\beta_2;t}= \sum_{a,b=R,L} \int_0^L\dd x_1\dd x_2\, \Psi_{\bbet}^{ab}(x_1,x_2;t) \psi_a^\dagger(x_1)\psi_b^\dagger(x_2)\ket0 ,
\label{eq:twostate}
\end{equation}
where
\begin{align}
 \Psi_{\bbet}^{ab}(x_1,x_2;t) ={}&\widehat f_{\beta_1}^{a}(x_1,t)\widehat f_{\beta_2}^{b}(x_2,t)\left[ A_{12}(\bbet)\theta(x_2-x_1) + A_{21}(\bbet)\theta(x_1-x_2) \right] \nonumber\\ &- \widehat f_{\beta_2}^{a}(x_1,t)\widehat f_{\beta_1}^{b}(x_2,t) \left[ A_{21}(\bbet)\theta(x_2-x_1) +A_{12}(\bbet)\theta(x_1-x_2)\right].
\label{eq:twoansatz}
\end{align}
This is the direct analogue of the standard coordinate Bethe-ansatz
form: the one-particle wavefunctions are multiplied by amplitudes
associated with the different orderings, and fermionic antisymmetry is
implemented explicitly. Equivalently,
\begin{equation}
 \Psi_{\bbet}^{ab}(x_1,x_2;t)=
 \begin{cases}
 A_{12}\widehat f_1^a(x_1)\widehat f_2^b(x_2) - A_{21}\widehat f_2^a(x_1)\widehat f_1^b(x_2), &x_1<x_2,\\[4pt]A_{21}\widehat f_1^a(x_1)\widehat f_2^b(x_2) - A_{12}\widehat f_2^a(x_1)\widehat f_1^b(x_2),&x_1>x_2,
 \end{cases}
\label{eq:twopiecewise}
\end{equation}
where the explicit time argument of the one-particle spinors has been
suppressed for notational simplicity. Away from $x_1=x_2$, each term in Eq.~\eqref{eq:twoansatz} is a product
of exact one-particle solutions and therefore satisfies the
time-dependent Schrodinger equation. The interaction imposes a
nontrivial condition only at the coincidence line.

\subsubsection{The two-particle S-matrix}

Let
\begin{equation}
 r=x_1-x_2.
\end{equation}
For the $RL$ component define, at coincidence,
\begin{equation}
 X=\widehat f_1^R\widehat f_2^L,\qquad Y=\widehat f_2^R\widehat f_1^L,
\end{equation}
and
\begin{equation}
 W=X-Y,\qquad D=X+Y.
\label{eq:WD}
\end{equation}
The values immediately on the two sides of the coincidence line are
\begin{align}
 \Psi^{RL}(0^-) &=A_{12}X-A_{21}Y,
 \nonumber\\
 \Psi^{RL}(0^+)
 &=A_{21}X-A_{12}Y.
\label{eq:contacttraces}
\end{align}
Consequently,
\begin{align}
 \Psi^{RL}(0^+)-\Psi^{RL}(0^-) &=(A_{21}-A_{12})D, \nonumber\\\Psi^{RL}(0^+)+\Psi^{RL}(0^-)&=(A_{12}+A_{21})W.
\label{eq:contactsumdiff}
\end{align}
Using the symmetric Bergknoff-Thacker prescription of
Eq.~\eqref{eq:prescription}, the $RL$ component of the two-particle
Schrodinger equation gives
\begin{equation}
 -2\ii\left[\Psi^{RL}(0^+)-\Psi^{RL}(0^-)\right]+g_0\left[ \Psi^{RL}(0^+)+\Psi^{RL}(0^-) \right]=0 .
\label{eq:contactjump}
\end{equation}
Substituting Eq.~\eqref{eq:contactsumdiff} yields
\begin{equation}
 -2\ii(A_{21}-A_{12})D +g_0(A_{12}+A_{21})W =0.
\label{eq:contacteq}
\end{equation}
Using the explicit one-particle solution in Eq.~\eqref{eq:fhat}, the two products at coincidence contain the same spacetime-dependent scalar phase.  Their sum and difference therefore retain only the relative rapidity weights, and one obtains
\begin{equation}
 \frac{W}{D} =\tanh\frac{\theta}{2}.
\label{eq:WDratio}
\end{equation}
Equation~\eqref{eq:contacteq} then gives
\begin{equation}
\frac{A_{21}(\beta_1,\beta_2)}{A_{12}(\beta_1,\beta_2)} = \frac{ 2\ii+g_0\tanh(\theta/2)}{2\ii-g_0\tanh(\theta/2)}.
\end{equation}
Using
\begin{equation}
 \cot\mu=-\frac{g_0}{2},
\label{eq:muagain}
\end{equation}
we obtain
\begin{equation}
 A_{21}(\beta_1,\beta_2)= \Smu(\theta) A_{12}(\beta_1,\beta_2),
\label{eq:Aexchange}
\end{equation}
where
\begin{align}
 \Smu(\theta)= \frac{1+\ii\cot\mu\tanh(\theta/2)} {1-\ii\cot\mu\tanh(\theta/2)}=-\frac{\sinh(\theta/2-\ii\mu)}{\sinh(\theta/2+\ii\mu)}.
\label{eq:Sdef}
\end{align}
The scalar S-matrix satisfies
\begin{equation}
 \Smu(\theta)\Smu(-\theta)=1.
\label{eq:unitarity}
\end{equation}
Thus the two-particle Schrodinger equation relates the amplitudes of
the two ordering sectors by the scalar S-matrix $\Smu(\theta)$. No additional space- or time-dependent factor appears
in this relation. We henceforth choose
\begin{equation}
 A(\beta_1,\beta_2)
 \equiv
 A_{12}(\beta_1,\beta_2)
\label{eq:referenceA}
\end{equation}
as the reference amplitude. The second ordering amplitude is then fixed
by
\begin{equation}
 A_{21}(\beta_1,\beta_2) = \Smu(\theta)A(\beta_1,\beta_2).
\label{eq:A21}
\end{equation}

\subsubsection{Periodic boundary conditions and the rapidity lattice}
\label{sec:periodicity}

The local two-particle solution above is not yet a periodic wavefunction.  In the spectrally normalized basis the one-particle winding relation is particularly simple,
\begin{equation}
 \widehat f_\beta(x+L,t)= \widehat f_{\beta-p}(x,t), \qquad p=\kappa L.
\label{eq:windingtwo}
\end{equation}
Thus a coordinate block with fixed rapidity is mapped to a neighboring point of the corresponding physical rapidity lattice.  The two-particle state must therefore contain the complete rapidity lattices
\begin{equation}
 \beta_j=\widetilde\beta_j+p n_j, \qquad n_j\in\mathbb Z,\qquad j=1,2 .
\label{eq:lattices}
\end{equation}
The constants $\widetilde\beta_j$ specify the origins of the two lattices.  They are arbitrary at the level of the exact formal solution and the difference equations determine only the relative amplitudes between neighboring lattice points. The periodic state is
\begin{equation}
 \ket{\Psi_2(t)}=\sum_{n_1,n_2\in\mathbb Z}\ket{\beta_1,\beta_2;t}, \qquad\beta_j=\widetilde\beta_j+p n_j.
\label{eq:twofullstate}
\end{equation}
Here each state $\ket{\beta_1,\beta_2;t}$ is built from the hatted one-particle spinors, with amplitudes $A(\beta_1,\beta_2)$ and $\Smu(\theta)A(\beta_1,\beta_2)$.  Periodicity is a property of the complete sum: a spatial winding maps one lattice point to another and is compensated by relabelling the corresponding integer summation variable.

\subsubsection{Transport of particle 1}

Consider first the periodic identification of $x_1$.  For $0<x_2<L$, the point $x_1=0$ lies in the ordering $x_1<x_2$, whereas $x_1=L$ lies in the ordering $x_1>x_2$.  At $x_1=L$, the term proportional to
\begin{equation}
 \widehat f_{\beta_1}^{a}(L,t) \widehat f_{\beta_2}^{b}(x_2,t)
\end{equation}
carries the amplitude $A_{21}(\beta_1,\beta_2)$.  Using Eq.~\eqref{eq:windingtwo},
\begin{equation}
 \widehat f_{\beta_1}^{a}(L,t)= \widehat f_{\beta_1-p}^{a}(0,t),
\end{equation}
this contribution is mapped directly to the rapidity block $(\beta_1-p,\beta_2)$.  In the term of the full lattice sum carrying these shifted rapidities, the same one-particle product at $x_1=0$ has coefficient $A(\beta_1-p,\beta_2)$.  Periodicity therefore requires
\begin{equation}
 A(\beta_1-p,\beta_2) = A_{21}(\beta_1,\beta_2),
\end{equation}
and hence
\begin{equation}
 A(\beta_1-p,\beta_2) =\Smu(\theta)A(\beta_1,\beta_2).
\label{eq:Ashift1}
\end{equation}
Since $\beta_1-p=\widetilde\beta_1+p(n_1-1)$, the shift is precisely the relabelling $n_1\to n_1-1$ in the complete rapidity lattice sum.

\subsubsection{Transport of particle 2}

Applying the same argument to the periodic identification of $x_2$ gives
\begin{equation}
 A_{21}(\beta_1,\beta_2-p) = A(\beta_1,\beta_2).
\label{eq:secondseamintermediate}
\end{equation}
The rapidity difference on the left-hand side is
\begin{equation}
 \beta_1-(\beta_2-p)=\theta+p.
\end{equation}
Using Eq.~\eqref{eq:A21} at the shifted rapidities therefore yields
\begin{equation}
 A(\beta_1,\beta_2-p) =\Smu(\theta+p)^{-1}A(\beta_1,\beta_2).
\label{eq:Ashift2}
\end{equation}
Again, $\beta_2-p=\widetilde\beta_2+p(n_2-1)$, so the winding is implemented by relabelling $n_2\to n_2-1$.

Equations~\eqref{eq:Ashift1} and \eqref{eq:Ashift2} are the scalar analytic difference equations for the reference amplitude in the hatted normalization.  Their especially simple form makes the separation of roles explicit: the one-particle dynamics and periodicity generate the rapidity lattices, while the interaction supplies only the scalar S-matrices that determine the relative weights on those lattices.

\subsubsection{Consistency of the two transports}

Define
\begin{equation}
 \mathcal K_1(\beta_1,\beta_2)=\Smu(\theta), \qquad \mathcal K_2(\beta_1,\beta_2)=\Smu(\theta+p)^{-1}.
\label{eq:twotransport}
\end{equation}
Then
\begin{equation}
 A(\bbet-p\be_j) = \mathcal K_j(\bbet)A(\bbet), \qquad j=1,2.
\end{equation}
Consistency requires
\begin{align}
\mathcal K_2(\beta_1-p,\beta_2)
 \mathcal K_1(\beta_1,\beta_2)=
 \mathcal K_1(\beta_1,\beta_2-p)
 \mathcal K_2(\beta_1,\beta_2).
\label{eq:twoflatness}
\end{align}
 This is the scalar discrete zero-curvature condition for the two-particle transport problem.  The shifted arguments remain essential even though the transport factors themselves are scalar. This condition has precisely the same form as the discrete zero-curvature condition associated with the qKZ equations \cite{PasnooriKondo2025,PasnooriQKZ2026}.  If
$Z_j(\boldsymbol{\beta})$ denotes the qKZ transport operator that shifts
the $j$th spectral variable by one step, qKZ consistency requires
\begin{equation}
 Z_j(\boldsymbol{\beta}-p\mathbf e_k)\, Z_k(\boldsymbol{\beta}) =Z_k(\boldsymbol{\beta}-p\mathbf e_j)\, Z_j(\boldsymbol{\beta}),
\end{equation}
which expresses the independence of the result from the order in which
the two discrete spectral translations are performed.  The distinction
is that in the previously studied qKZ problems the $Z_j$ are
matrix-valued operators acting in an internal spin space, whereas in
the present massive-Thirring construction the corresponding transport
is scalar and acts directly on the physical rapidity lattice.
\subsubsection{Static Bethe limit and the plane-wave normalization}
\label{sec:staticlimit}

The hatted normalization is adapted to finite-step periodic transport.  It makes the rapidity lattice explicit, but the phase in Eq.~\eqref{eq:fhat} is rapidly varying as $\kappa\to0$.  To make contact with the conventional static Bethe ansatz, we now introduce a second normalization only for this limit.  Define
\begin{equation}
 a(\beta)=e^{-\ii m_*\cosh\beta/\kappa},\qquad f_\beta(x,t)=a(\beta)^{-1}\widehat f_\beta(x,t).
\label{eq:spectralnormalization}
\end{equation}
Equivalently, $\widehat f_\beta=a(\beta)f_\beta$.  Using Eq.~\eqref{eq:fhat},
\begin{equation}
 \begin{pmatrix}f_\beta^R\\ f_\beta^L\end{pmatrix} =e^{\kappa t/2}
 \begin{pmatrix}e^{(\beta-\kappa x)/2}\\-\ii e^{-(\beta-\kappa x)/2}\end{pmatrix}
 \exp\!\left[-\frac{\ii}{\kappa} \{\mb(t)\cosh(\beta-\kappa x)-m_*\cosh\beta\}\right].
\label{eq:fexplicit}
\end{equation}
At fixed $x,t,\beta$ this normalization has the smooth limit
\begin{equation}
 f_\beta(x,t)\underset{\kappa\to0}{\longrightarrow}
 \begin{pmatrix}e^{\beta/2}\\-\ii e^{-\beta/2}\end{pmatrix}
 e^{\ii m_*x\sinh\beta-\ii m_*t\cosh\beta}.
\label{eq:fstatic}
\end{equation}
Thus $f_\beta$ is the normalization in which the ordinary massive Dirac plane wave and its propagation phase are manifest. The pure-shift winding relation \eqref{eq:hatwinding} becomes, in this normalization,
\begin{align}
 f_\beta(x+L,t)=P(\beta)f_{\beta-p}(x,t),
 \label{eq:winding}\end{align}
 where
 \begin{align}
 P(\beta)=\frac{a(\beta-p)}{a(\beta)}=\exp\!\left[\frac{2\ii m_*}{\kappa} \sinh\frac p2\sinh\!\left(\beta-\frac p2\right)\right].
\label{eq:P}
\end{align}
Equivalently,
\begin{equation}
 a(\beta-p)=P(\beta)a(\beta).
\label{eq:arecurrence}
\end{equation}
The hatted and plane-wave normalizations describe the same coordinate state; they merely place the one-particle rapidity phase in different factors.

Let $A^{(f)}$ denote the two-particle reference amplitude when the elementary spinors are written in the plane-wave normalization.  Equality of the complete coordinate block gives
\begin{equation}
 A^{(f)}(\beta_1,\beta_2) =a(\beta_1)a(\beta_2)A(\beta_1,\beta_2).
\label{eq:Afdef}
\end{equation}
Using Eq.~\eqref{eq:arecurrence} together with Eqs.~\eqref{eq:Ashift1} and \eqref{eq:Ashift2}, the finite-step transport equations become
\begin{align}
 A^{(f)}(\beta_1-p,\beta_2) &=P(\beta_1)\Smu(\theta)A^{(f)}(\beta_1,\beta_2),\nonumber\\A^{(f)}(\beta_1,\beta_2-p)&=P(\beta_2)\Smu(\theta+p)^{-1} A^{(f)}(\beta_1,\beta_2).
\label{eq:Aftransport}
\end{align}
Now the limit $p=\kappa L\to0$ is transparent because
\begin{equation}
 P(\beta)\longrightarrow e^{\ii m_*L\sinh\beta}.
\end{equation}
The two transport equations reduce to
\begin{equation}
 \Smu(\theta)e^{\ii k_1L}=1,\qquad e^{\ii k_2L}=\Smu(\theta),\qquad k_j=m_*\sinh\beta_j,
\label{eq:staticseam}
\end{equation}
which are the usual two-particle periodic Bethe conditions.  For $N$ particles the same argument gives
\begin{equation}
 e^{\ii m_*L\sinh\beta_j} =\prod_{k\neq j} \Smu(\beta_j-\beta_k)^{-1}.
\label{eq:staticBE}
\end{equation}
Thus the two normalizations serve complementary purposes: $\widehat f_\beta$ exposes the finite-step rapidity translation and the resulting physical rapidity lattice, while $f_\beta$ is introduced only here to display the conventional static plane-wave propagation phase and recover the standard Bethe equations in the static limit.

\section{Analytic difference equations and the explicit wavefunction}
\label{sec:solution}

The local scattering relations together with periodicity reduce the
many-body problem to a set of scalar analytic difference equations for
the amplitudes defined on the physical rapidity lattice.  A finite
shift of one rapidity by the lattice spacing relates the amplitude at
one lattice point to that at the neighboring point through the
appropriate massive-Thirring scalar S-matrix.  In this section we
solve these difference equations explicitly and use their solutions to
construct the complete coordinate-space wavefunction.  We then derive
a complementary momentum-space representation of the one-particle
solution, in which the time-dependent Dirac equation reduces to a
Whittaker equation, and discuss how the resulting analytic-difference
structure is related to that appearing in qKZ systems.

\subsection{The scalar recurrence and its \texorpdfstring{$q$}{q}-gamma solution}

In the hatted normalization the one-particle winding phase has already been absorbed into the elementary spinors.  The reference amplitude therefore contains only the interaction-dependent weight.  Up to an overall normalization associated with the choice of rapidity-lattice origins, we write
\begin{equation}
 A(\beta_1,\beta_2)=\Phi_p(\theta),\qquad \theta=\beta_1-\beta_2.
\end{equation}
Equation~\eqref{eq:Ashift1} then becomes the single first-order recurrence
\begin{equation}
 \Phi_p(\theta-p)=\Smu(\theta)\Phi_p(\theta).
\label{eq:phirecurrence}
\end{equation}
The second transport equation follows from the same recurrence evaluated at $\theta+p$ and unitarity.
For $p>0$, introduce
\begin{equation}
 q=e^{-p},\qquad y=e^\theta,\qquad  r=e^{\ii(\pi-2\mu)},\qquad \log r=\ii(\pi-2\mu).
\label{eq:qr}
\end{equation}
Then
\begin{equation}
 \Smu(\theta)=r^{-1}\frac{1+ry}{1+r^{-1}y}.
\end{equation}
Using the $q$-Pochhammer product, the solution is
\begin{equation}
\displaystyle
 \Phi_p(\theta)= e^{(\log r)\theta/p} \frac{(-r^{-1}e^\theta;q)_\infty}{(-r e^\theta;q)_\infty}.
\label{eq:phiproduct}
\end{equation}
The identity $(aq;q)_\infty=(a;q)_\infty/(1-a)$ verifies Eq.~\eqref{eq:phirecurrence} directly. A constant normalization may be chosen, for example at $\theta=0$. To express the same solution in terms of $q$-gamma functions, define
\begin{equation}
 \eta_q=\frac{\log r}{\log q},\qquad\zeta=\frac{\theta+\ii\pi}{\log q},\qquad q^{\eta_q}=r,\qquad q^\zeta=-e^\theta.
\end{equation}
With
\begin{equation}
 \Gamma_q(z)=(1-q)^{1-z}\frac{(q;q)_\infty}{(q^z;q)_\infty},
\end{equation}
we obtain
\begin{equation}
\displaystyle
 \Phi_p(\theta)=e^{-\eta_q\theta}(1-q)^{2\eta_q}\frac{\Gamma_q(\zeta+\eta_q)}{\Gamma_q(\zeta-\eta_q)}.
\label{eq:phigamma}
\end{equation}
These are standard $q$-special-function identities \cite{GasperRahman2004,DLMF}. The limit $p\to0$ is nonuniform in these variables because both $\eta_q$ and $\zeta$ scale as $1/p$. For $p<0$, one uses the inverse recurrence and products with $q^{-1}$, or an appropriate analytic continuation. We use $p>0$ in the explicit product formulas and keep the direction of the physical drive explicit in the discussion of future dynamical applications. The limit $p\to0$ is nonuniform at the level of the individual $q$-special functions.  The conventional static limit is therefore taken at the level of the transport equations, as described in Sec.~\ref{sec:staticlimit}, rather than by assuming pointwise convergence of Eq.~\eqref{eq:phiproduct}.

The solution of the analytic difference equation is not unique in the
usual functional sense.  If $\Phi_p(\theta)$ satisfies (\ref{eq:phirecurrence})
then so does \begin{equation}
 h(\theta)\Phi_p(\theta), \qquad h(\theta-p)=h(\theta).
\end{equation}
This is the familiar periodic-multiplier ambiguity of first-order
analytic difference equations, and in the mathematical treatment of
Ruijsenaars it is removed by imposing additional analyticity and
growth conditions that select a distinguished minimal solution.  In
the present problem, however, the physical rapidity-lattice
construction makes the ambiguity considerably less severe.  On a
fixed lattice
\begin{equation}
 \theta_n=\widetilde\theta+np ,
\end{equation}
any $p$-periodic function is constant,
$h(\theta_n)=h(\widetilde\theta)$, and therefore multiplies all
amplitudes on that lattice by the same overall factor.  It does not
modify the relative amplitudes between neighboring lattice points,
which remain fixed uniquely by the massive-Thirring scalar S-matrix.
More generally, in the $N$-particle state a periodic multiplier depends
only on the continuous lattice origins $\widetilde\beta_i$ and not on
the discrete lattice labels.  For fixed origins it can therefore be
absorbed into the normalization of the state.  Only if one later forms
superpositions involving different choices of lattice origins does an
additional criterion, such as minimal analyticity or an asymptotic
normalization condition, become necessary to fix this residual
freedom.
\subsection{The \texorpdfstring{$N$}{N}-particle sector}
\label{sec:N}

The two-particle construction generalizes directly to arbitrary particle number.  We continue to use the spectrally normalized one-particle blocks $\widehat f_\beta$, so that spatial winding acts as a pure rapidity translation.  Let $Q\in S_N$ denote a permutation of the particle labels and define
\begin{equation}
 \Theta_Q(\bx)= \theta(x_{Q(2)}-x_{Q(1)}) \theta(x_{Q(3)}-x_{Q(2)})\cdots \theta(x_{Q(N)}-x_{Q(N-1)}),
\end{equation}
so that $\Theta_Q$ is nonzero for the ordering 
\begin{equation}
 x_{Q(1)}<x_{Q(2)}<\cdots<x_{Q(N)}.
\end{equation}
The $N$-particle coordinate wavefunction is
\begin{equation}
 \Psi_{\bbet}^{a_1\cdots a_N}(\bx,t) = \mathcal A \left[ \prod_{j=1}^{N}\widehat f_{\beta_j}^{a_j}(x_j,t) \sum_{Q\in S_N} A_Q(\bbet)\Theta_Q(\bx) \right],
\label{eq:Nansatz}
\end{equation}
where $\mathcal A$ antisymmetrizes the particle labels and
\begin{equation}
 \bbet=(\beta_1,\ldots,\beta_N).
\end{equation}
The quantities $A_Q(\bbet)$ are the amplitudes corresponding to the orderings labelled by $Q$ and are defined in the hatted normalization. Away from the coincidence hyperplanes $x_i=x_j$, every term in Eq.~\eqref{eq:Nansatz} is a product of exact one-particle solutions and therefore satisfies the time-dependent Schrodinger equation.  At a generic coincidence of two neighboring particles, all remaining factors are smooth.  The two-particle S-matrix calculation therefore applies unchanged, and two ordering sectors that differ by exchanging adjacent particles $i$ and $j$ obey
\begin{equation}
 A_{\ldots ji\ldots}(\bbet) = \Smu(\beta_i-\beta_j) A_{\ldots ij\ldots}(\bbet).
\label{eq:AQexchange}
\end{equation}
Thus all amplitudes corresponding to different orderings are determined by one amplitude of our choosing.  We take the reference amplitude corresponding to the order
\begin{equation}
 x_1<x_2<\cdots<x_N
\end{equation}
and denote its amplitude by
\begin{equation}
 A_N(\bbet)\equiv A_{12\cdots N}(\bbet).
\label{eq:referenceAN}
\end{equation}
Since the S-matrix is scalar, different sequences of adjacent exchanges give the same result; the many-body factorization condition reduces to the unitarity relation $\Smu(\theta)\Smu(-\theta)=1$. The remaining reference amplitude is fixed by periodic transport.  To see this consider transporting particle $j$ once around the system.  Starting from the reference ordering, the particle first passes $j+1,\ldots,N$, contributing
\begin{equation}
 \prod_{k>j}\Smu(\beta_j-\beta_k).
\end{equation}
It then crosses the periodic seam.  In the hatted normalization there is no additional scalar winding phase; Eq.~\eqref{eq:hatwinding} simply gives
\begin{equation}
 \beta_j\longrightarrow\beta_j-p, \qquad p=\kappa L.
\end{equation}
To restore the reference ordering, the shifted particle passes $1,\ldots,j-1$, producing
\begin{equation}
 \prod_{k<j}\Smu(\beta_j-\beta_k-p).
\end{equation}
Hence
\begin{equation}
\displaystyle
 A_N(\bbet-p\be_j) = \mathcal K_j(\bbet)A_N(\bbet),
\label{eq:Ntransport}
\end{equation}
with
\begin{equation}
\displaystyle\mathcal K_j(\bbet)=\prod_{k<j}\Smu(\beta_j-\beta_k-p) \prod_{k>j}\Smu(\beta_j-\beta_k).
\label{eq:Ktransport}
\end{equation}
This form cleanly separates the two ingredients of the construction.  The one-particle equation and periodicity generate the shift $\beta_j\to\beta_j-p$, whereas the interaction determines the transport weight $\mathcal K_j$ through two-particle scattering. The transport equations are mutually compatible.  For $i\neq j$,
\begin{equation}
 \mathcal K_i(\bbet-p\be_j)\mathcal K_j(\bbet) =\mathcal K_j(\bbet-p\be_i)\mathcal K_i(\bbet).
\label{eq:flatness}
\end{equation}
For example, when $i<j$, all factors not involving the pair $(i,j)$ cancel, while the remaining terms reduce to
\begin{equation}
 \Smu(\theta+p)\Smu(-\theta-p) = \Smu(\theta)\Smu(-\theta) =1.
\end{equation}
Equation~\eqref{eq:flatness} is the scalar discrete zero-curvature condition associated with periodic transport. The pairwise function obtained above obeys
\begin{equation}
 \Phi_p(\theta-p) = \Smu(\theta)\Phi_p(\theta).
\label{eq:phirecurrence_recalled}
\end{equation}
It follows directly that an explicit solution of the complete $N$-particle transport system in the hatted normalization is
\begin{equation}
 A_N(\boldsymbol\beta) =\prod_{1\leq i<j\leq N} \Phi_p(\beta_i-\beta_j).
\label{eq:AN}
\end{equation}
Indeed, shifting $\beta_j$ by $-p$ produces the factors in Eq.~\eqref{eq:Ktransport}: pairs with $j<k$ use Eq.~\eqref{eq:phirecurrence_recalled} directly, while pairs with $k<j$ use the same recurrence at the shifted argument together with unitarity. Using the $q$-gamma representation,
\begin{equation}
 \Phi_p(\theta) =e^{-\eta_q\theta}(1-q)^{2\eta_q}\frac{\Gamma_q(\zeta+\eta_q)} {\Gamma_q(\zeta-\eta_q)},\qquad q=e^{-p},
\end{equation}
with
\begin{equation}
 \zeta_{ij}= \frac{\beta_i-\beta_j+\ii\pi}{\log q},
\end{equation}
the reference amplitude becomes
\begin{equation}
 A_N(\boldsymbol\beta) = \prod_{1\leq i<j\leq N}\left[ e^{-\eta_q(\beta_i-\beta_j)} (1-q)^{2\eta_q} \frac{\Gamma_q(\zeta_{ij}+\eta_q)}{\Gamma_q(\zeta_{ij}-\eta_q)}\right].
\label{eq:explicitAN}
\end{equation}
The absence of a separate one-particle factor in Eqs.~\eqref{eq:AN} and \eqref{eq:explicitAN} is a consequence of the normalization: that factor has been absorbed into $\widehat f_\beta$.

A fixed set of rapidities does not by itself define a periodic state because spatial winding moves one rapidity to a neighboring point of its physical rapidity lattice.  We therefore introduce
\begin{equation}
 \beta_j=\widetilde\beta_j+p n_j, \qquad n_j\in\mathbb Z, \qquad j=1,\ldots,N,
\label{eq:Nlattices}
\end{equation}
and sum over the complete set of rapidity lattices.  The periodic $N$-particle wavefunction is
\begin{align}
 \ket{\Psi_N(t)} ={}& \sum_{\bn\in\mathbb Z^N}\sum_{a_1,\ldots,a_N}\int_0^L\prod_{j=1}^{N}\dd x_j\, \psi_{a_1}^\dagger(x_1)\cdots \psi_{a_N}^\dagger(x_N)
 \nonumber\\&\times\mathcal A \left[ \prod_{j=1}^{N} \widehat f_{\beta_j}^{a_j}(x_j,t)\sum_{Q\in S_N}A_Q(\bbet)\Theta_Q(\bx) \right]\ket0, \qquad \bbet=\widetilde{\bbet}+p\bn .
\label{eq:fullstate}
\end{align}
Every $A_Q$ is obtained from the reference amplitude $A_N$ by successive applications of Eq.~\eqref{eq:AQexchange}.

The distinction between the reference amplitude and the complete state is important.  The analytic difference equations determine the relative weights of neighboring points of the rapidity lattice, but periodicity belongs to the complete rapidity lattice sum.  Under $x_j\to x_j+L$, Eq.~\eqref{eq:hatwinding} maps the corresponding elementary spinor from $\beta_j$ to $\beta_j-p$.  Since
\begin{equation}
 \beta_j-p = \widetilde\beta_j+p(n_j-1),
\end{equation}
the transformed contribution is already present in Eq.~\eqref{eq:fullstate}.  Relabelling $n_j\to n_j-1$ and using Eq.~\eqref{eq:Ntransport} leaves the complete wavefunction invariant.

The sum in Eq.~\eqref{eq:fullstate} should be understood as the \textit{Jackson-type sum} associated with the analytic difference equations.  At this stage it is a formal representation of the exact solution.  The rapidity lattices themselves originate from the one-particle winding law, while the difference equations determine the interaction-dependent weights carried by neighboring lattice sectors.  We do not impose an additional convergence prescription on this formal sum.  Its realization as a physical many-body state depends on the state under consideration and on the asymptotic regime in which physical observables are extracted.  This is analogous to Jackson-type sums appearing in off-shell qKZ solutions, although here the lattice variables are the physical particle rapidities rather than auxiliary Bethe rapidities.

The constants $\widetilde\beta_j$ specify the origins of the rapidity lattices.  They are not fixed by the analytic difference equations, which determine only relative amplitudes along each rapidity lattice.  As in Jackson-type difference-equation solutions, they are therefore left as arbitrary constants at the level of the exact formal state.  Their role in selecting physical states belongs to the subsequent asymptotic Bethe ansatz analysis, such as in \cite{PasnooriTransmutation}, and will not be addressed here.

\subsection{Momentum representation and Whittaker equations}
\label{sec:Whittaker}

The scalar difference equations determine how neighboring rapidities are combined in the many-body state. There is also a complementary differential description of the one-particle propagation. Because the bare mass is spatially homogeneous, write
\begin{equation}
 f^{R,L}(x,t)=e^{\ii kx}(R_k(t),L_k(t)).
\end{equation}
The component equations are
\begin{equation}
 \dot R_k=-\ii kR_k+\mb L_k,\qquad \dot L_k=\ii kL_k-\mb R_k.
\label{eq:modeDirac}
\end{equation}
The Fourier decomposition is valid, although the rapidity basis of Section~\ref{sec:one} is better adapted to the coordinate contact condition. Introducing
\begin{equation}
 \chi_+=R_k-\ii L_k,\qquad \chi_-=R_k+\ii L_k,
\end{equation}
we obtain
\begin{align}
 \dot\chi_+=\ii\mb\chi_+-\ii k\chi_-,&\;\;\;\;\;\;\;\;
 \dot\chi_-=-\ii k\chi_+-\ii\mb\chi_-,
 \label{eq:chifirst}\\
 \ddot\chi_\pm+[k^2+&\mb^2\mp\ii\dot\mb]\chi_\pm=0.
\label{eq:chipartners}
\end{align}
For the exponential mass, define
\begin{equation}
 y=\frac{2\ii\mb(t)}\kappa,\qquad \chi_\pm=y^{-1/2}\mathcal W_\pm(y).
\label{eq:Whittakerchange}
\end{equation}
Equation~\eqref{eq:chipartners} becomes
\begin{equation}
 \mathcal W_\pm''+\left[-\frac14+\frac{a_\pm}{y} +\frac{1/4-b_{\mathrm W}^2}{y^2}\right]\mathcal W_\pm=0, \qquad a_\pm=\mp\frac12,\qquad b_{\mathrm W}=\frac{\ii k}{\kappa}.
\label{eq:Whittaker}
\end{equation}
This is Whittaker's equation. Its standard solutions are the confluent hypergeometric functions $M_{a,b}(y)$ and $W_{a,b}(y)$ in Whittaker notation \cite{DLMF}. For generic parameters, one may choose
\begin{equation}
 \chi_+(t)=y^{-1/2}\left[
 A_k M_{-1/2,b_{\mathrm W}}(y)+B_k W_{-1/2,b_{\mathrm W}}(y)\right],
\label{eq:Whittakersolution}
\end{equation}
and obtain the other component from
\begin{equation}
 \chi_- =\frac{\mb\chi_++\ii\dot\chi_+}{k},\qquad k\ne0.
\label{eq:chireconstruct}
\end{equation}
At exceptional parameters one uses an independent limiting basis of the same equation. The two second-order solutions are not independent spinors: Eq.~\eqref{eq:chifirst} leaves only two integration constants for a fixed $k$. For $k=0$, the solutions are elementary,
\begin{equation}
 \chi_\pm(t)=C_\pm \exp\!\left[\pm\ii\int_{t_0}^{t}\mb(s)\dd s\right].
\end{equation}
The original components follow from $R_k=(\chi_++\chi_-)/2$ and $L_k=(\chi_-\chi_+)/(2\ii)$. The connection to the rapidity representation can be seen directly.
Consider the spectrally normalized block $\widehat f$ of
Eq.~\eqref{eq:fhat} and the integral
\begin{equation}
 \mathcal F_k(x,t)=\int_{\mathcal C}\dd\beta\, e^{\ii k\beta/\kappa}\widehat f_\beta(x,t),
\label{eq:Fourierbeta}
\end{equation}
where the transform is first defined in a domain in which the integral
converges and is subsequently continued analytically to real time.
The change of variables $v=\beta-\kappa x$ extracts the spatial factor
$e^{\ii kx}$.  Using
\begin{equation}
 \int_{-\infty}^{\infty}\dd v\,e^{-z\cosh v+\nu v} = 2K_\nu(z), \qquad \operatorname{Re}z>0,
\label{eq:BesselIntegral}
\end{equation}
where $K_\nu$ is the modified Bessel function. Analytically continuing to $z=\ii\mb(t)/\kappa$, one obtains

\begin{equation}
 \mathcal F_k(x,t)=2e^{\ii kx}e^{\kappa t/2}
 \begin{pmatrix}
 K_{1/2+\ii k/\kappa}(\ii\mb/\kappa)\\[2pt] -\ii K_{-1/2+\ii k/\kappa}(\ii\mb/\kappa)
 \end{pmatrix}.
\label{eq:Besselmodes}
\end{equation}
Analytic continuation specifies the real-time branch \cite{DLMF}. Its spinor combinations satisfy Eq.~\eqref{eq:Whittaker}, so this integral is an explicit transform between the null-coordinate family and a momentum-resolved Whittaker solution.

For one particle, Poisson summation of the periodic rapidity lattice makes the relation particularly transparent:
\begin{equation}
 \sum_{n\in\mathbb Z}\widehat f_{\widetilde\beta+pn}(x,t) =\frac1p\sum_{\ell\in\mathbb Z} e^{-2\pi\ii\ell\widetilde\beta/p}\mathcal F_{k_\ell}(x,t), \qquad k_\ell=\frac{2\pi\ell}{L}.
\label{eq:onePoisson}
\end{equation}
The spatial Fourier momenta therefore emerge from the rapidity step. Equation \eqref{eq:onePoisson} is the Poisson representation of the formal one-particle rapidity-lattice sum and makes explicit the corresponding discrete momentum modes on the circle and it does not require a small-step limit. In the many-body wavefunction, the pairwise $q$-gamma weights couple these elementary blocks.

\subsection{Bosonization, qKZ and analytic difference equations:
related mathematical structures}
\label{sec:mathcontext}

The exact constructions discussed in this paper arise from field
theories that are physically related, rather than from unrelated models
whose difference equations happen to look similar.  This observation
provides the natural starting point for comparing their mathematical
structures. The $U(1)$-symmetric Thirring model separates, under non-Abelian
bosonization, into a free charge sector and an interacting spin sector
described by an anisotropic current-current deformation of the
$SU(2)_1$ WZW model.  Abelian bosonization of the latter gives the
sine-Gordon model, while refermionization gives the massive
Thirring model in the corresponding massive regime.  The isotropic
limit of the current-current theory gives the $SU(2)$ Gross-Neveu
model.  These relations are summarized in Fig.~(\ref{fig:picture1}). These equivalent descriptions organize integrability in different
ways.  In the generalized Bethe ansatz for the $SU(2)$ Gross-Neveu
model, periodic transport produces rational qKZ equations associated
with the XXX $R$-matrix.  In the anisotropic $U(1)$-symmetric Thirring
model the corresponding construction produces trigonometric or hyperbolic qKZ
equations associated with the XXZ $R$-matrix.  Their amplitudes are
vector valued in the internal spin space, and the off-shell Bethe
ansatz introduces auxiliary rapidities together with scalar difference
equations for their weights.  In the present massive-Thirring
description there is no independent nested spin space.  Instead,
periodic transport acts directly on the physical rapidities, producing
the scalar analytic difference equations derived above. The relation between these structures is therefore deeper than the
appearance of similar scattering phases.  Bosonization and
refermionization identify the underlying interacting spin degrees of freedom of the chiral theories with the bare fermions of the massive Thirring model, so it
is natural to expect the algebraic descriptions of its integrability to
be related as well.  Part of this relation is already visible in the
existing mathematical literature.

Quantum affine algebras provide the first bridge.  Matrix elements of
quantum-affine intertwining operators satisfy qKZ equations
\cite{FrenkelReshetikhin1992}.  For \(U_q(\widehat{\mathfrak{sl}}_2)\), the quantum affine algebra underlying the XXZ \(R\)-matrix and its associated qKZ equations, these intertwining operators admit free-field realizations. The resulting integral representations were shown by Kuroki and Nakayashiki to coincide with the hypergeometric qKZ solutions of Tarasov and Varchenko \cite{KurokiNakayashiki2008}.  On the sine-Gordon/Thirring side,
nonlocal conserved charges generate quantum-affine symmetries that
constrain the factorized scattering
\cite{BernardLeClair1991}.  Sine-Gordon form-factor equations also admit
a level-zero qKZ formulation
\cite{JimboKojimaMiwaQuano1994}, while free-field constructions provide
bosonic realizations of the Zamolodchikov-Faddeev algebra and exact
form factors \cite{Lukyanov1993}.  Thus qKZ equations, quantum-affine
intertwiners, free fields and sine-Gordon/Thirring integrability are
already parts of a common algebraic framework.  What remains open in
the present problem is the precise place of the physical rapidity
lattice generated by the time-dependent massive-Thirring wavefunction
within this framework.

A distinctive feature of the coordinate construction is that the rapidity lattice is forced already by the one-particle problem.  The time-dependent mass fixes the rapidity dependence of the elementary spinor, so a fixed-rapidity block cannot be made periodic by choosing an arbitrary chiral function.  In the spectrally normalized basis, spatial winding acts simply as
\begin{equation}
 \widehat f_{\beta_j}(x_j+L,t) = \widehat f_{\beta_j-p}(x_j,t), \qquad p=\kappa L.
\end{equation}
The complete periodic state must therefore contain
\begin{equation}
 \beta_j=\widetilde\beta_j+p n_j, \qquad n_j\in\mathbb Z .
\label{eq:mathcontextlattice}
\end{equation}
This lattice exists before the interaction-dependent difference equation is solved.  The role of the interaction is instead to determine the relative weights carried by neighboring lattice points.  In the two-particle sector the corresponding pairwise weight obeys
\begin{equation}
 \Phi_p(\theta-p) =\Smu(\theta)\Phi_p(\theta).
\label{eq:RuijsenaarsADE}
\end{equation}
This belongs to the broad class of first-order analytic difference equations studied by Ruijsenaars \cite{Ruijsenaars1997}; the appearance of generalized gamma functions, and here of $q$-gamma functions, is natural from that viewpoint. The complete massive-Thirring state nevertheless contains substantially more information than Eq.~\eqref{eq:RuijsenaarsADE}.  Periodicity generates the physical rapidity lattices, the generalized coordinate Bethe ansatz supplies the ordering sectors, and the particle-particle interaction leads to the scalar S-matrix which fixes the relative weights that enter the recurrence.  It is therefore useful to keep this logical hierarchy conceptually separate from a first-order difference equation considered in isolation.  

There is also a second, mathematically precise appearance of
Ruijsenaars systems on the qKZ side.  Cherednik's affine-Hecke
formulation of qKZ leads, through the Cherednik-Matsuo correspondence,
to scalar joint eigenfunctions of commuting Cherednik-Macdonald
difference operators \cite{Cherednik1992,Stokman2011}.  In type $A$
these are closely related to trigonometric
Macdonald-Ruijsenaars operators.  Double affine Hecke algebras (DAHA) provide
a bispectral formulation of this correspondence
\cite{vanMeerStokman2010}, and an $R$-matrix version of the
qKZ-Ruijsenaars relation was developed by Zabrodin and Zotov
\cite{ZabrodinZotov2017}. It is important to note that the
Cherednik-Matsuo map is a concrete correspondence between qKZ
solutions and a Macdonald-Ruijsenaars spectral problem.  By contrast,
Eq.~\eqref{eq:RuijsenaarsADE} is simply a first-order
Ruijsenaars-type analytic difference equation arising directly from
massive-Thirring periodicity.  We do not identify $\Phi_p$ with a
Macdonald-Ruijsenaars eigenfunction.  The physical bosonization
relation gives an independent reason to expect a deeper connection
between the two structures, but establishing that connection is a
separate mathematical problem.

Several neighboring structures make this comparison particularly
interesting.  Baxter $Q$-operators have been constructed for quantum
Ruijsenaars systems as commuting integral operators associated with
Macdonald-Ruijsenaars Hamiltonians
\cite{BelousovDerkachovKharchevKhoroshkin2024}.  This does not make
Eq.~\eqref{eq:RuijsenaarsADE} a Baxter $TQ$ equation, nor does it
identify $\Phi_p$ with a $Q$-function.  It does, however, suggest asking
whether the pairwise weight or the complete rapidity-lattice state has
a useful representation in terms of a Baxter-type integral operator. Whittaker structures provide another nearby connection.  Degenerations
of Macdonald theory lead to $q$-Toda systems and $q$-Whittaker
functions; Cherednik and collaborators developed both symmetric and
nonsymmetric versions using DAHA and nil-DAHA methods
\cite{Cherednik2009,CherednikOrr2015}. The ordinary Whittaker functions obtained in the present paper arise
differently: resolving the one-particle massive-Thirring equation in
spatial momentum gives the ordinary one-variable Whittaker differential
equation for the exponential mass protocol.  These functions should
not be identified with the root-system $q$-Whittaker functions above.
Likewise, the quantity $q=e^{-\kappa L}$ appearing in the present
$q$-gamma solution measures the physical winding-induced rapidity step
and is not automatically a quantum-group or Macdonald deformation
parameter.

\begin{center}
\begin{figure}
\includegraphics[width=0.98\columnwidth]{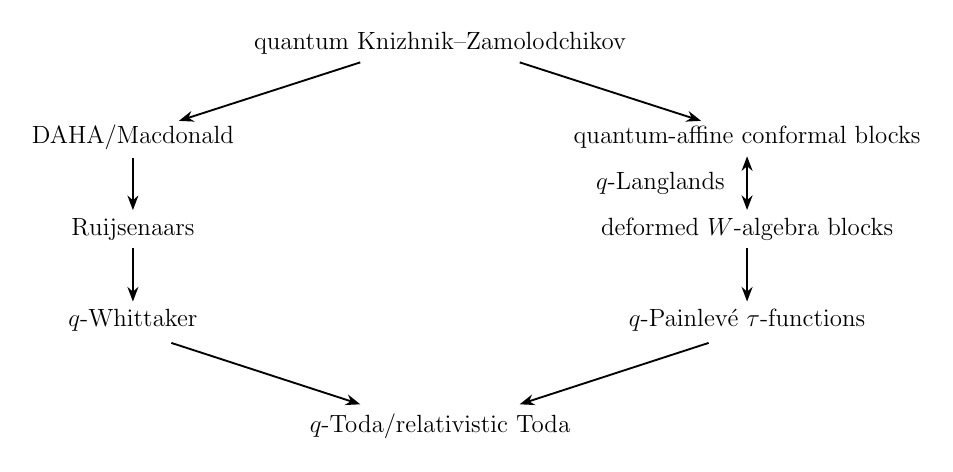}
\caption{Schematic network of mathematical structures related to qKZ equations.
One branch connects qKZ to DAHA/Macdonald theory, Ruijsenaars systems,
and $q$-Whittaker functions, while a second proceeds through
quantum-affine conformal blocks, $q$-Langlands duality, deformed
$W$-algebra blocks, and $q$-Painlev\'e $\tau$-functions.  The two
branches meet naturally in the $q$-Toda/relativistic-Toda framework}
\label{fig:picture2}
\end{figure}
\end{center}

Quantum $K$-theory gives a complementary mathematical realization.
Koroteev and Zeitlin showed that quantum $K$-theoretic vertex functions
can simultaneously participate in qKZ structures and solve
trigonometric Ruijsenaars-Schneider spectral problems
\cite{KoroteevZeitlin2021}.  In a related direction, the quantum
$q$-Langlands correspondence of Aganagic, Frenkel and Okounkov relates
$q$-deformed conformal blocks of quantum affine algebras to conformal
blocks of deformed $W$-algebras, with quantum $K$-theory of Nakajima
quiver varieties providing the geometric setting
\cite{AganagicFrenkelOkounkov2018}.  Since qKZ equations arise as
difference equations for quantum-affine conformal blocks, this provides
a route from the qKZ side of the present discussion to deformed
$W$-algebra structures. These relations are schematically represented by the left column of Fig.~(\ref{fig:picture2})

It is also interesting to note that Painlev\'e theory enters through this same conformal-block and
gauge-theory branch rather than as an independent analogy.  In the
$A_1$ case, quantum $q$-Langlands relates quantum-affine blocks to
$q$-Virasoro blocks, while $q$-Painlev\'e tau functions admit
expansions in $q$-Virasoro Whittaker blocks, equivalently in
five-dimensional Nekrasov partition functions
\cite{BershteinShchechkin2017}. 
Nekrasov partition functions and quantum $K$-theory provide the
gauge-theory realization of the intermediate objects.  Painlev\'e
theory is therefore connected to the same qKZ-quantum-affine network
already relevant to the generalized Bethe ansatz, rather than directly
to the first-order massive-Thirring recurrence. These relations are schematically represented by the right column of Fig.~(\ref{fig:picture2}).
The two branches above are not completely independent.  In the
rank-one case they meet naturally through the relativistic Toda
system.  The $q$-Whittaker functions arising as degenerations of
Macdonald-Ruijsenaars theory are eigenfunctions of $q$-Toda
difference Hamiltonians \cite{Etingof1999}.  On the other hand, the
two-particle affine relativistic Toda system provides the autonomous
integrable system whose deautonomization gives the
$q$-Painlev\'e equation of type $A_7^{(1)\prime}$
\cite{BershteinGavrylenkoMarshakov2018}.  The same relativistic-Toda
spectral curve is the Seiberg-Witten curve of pure five-dimensional
$SU(2)$ gauge theory.  In this gauge-theory realization the
$q$-Painlev\'e tau function is expressed as a Fourier-type sum of
$q$-Virasoro Whittaker conformal blocks, equivalently of five-dimensional
Nekrasov partition functions \cite{BershteinShchechkin2017}. The $q$-Whittaker functions appearing on the left
and the $q$-Virasoro Whittaker conformal blocks entering the
$q$-Painlev\'e tau function are distinct mathematical objects.  Their
relation here is through the common relativistic-Toda and
five-dimensional gauge-theory framework rather than through a direct
identification. All these connections are summarized in Fig.~(\ref{fig:picture2}).

A conceptually different, but potentially complementary, gauge-theoretic
perspective is provided by four-dimensional Chern-Simons theory.  In
the Costello-Witten-Yamazaki construction, Yang-Baxter structures,
$R$-matrices and integrable field theories arise directly from a
four-dimensional gauge theory \cite{Costello1,Costello2,costello3,Yamazaki2025Overview}.  Recent work
by Komatsu, Sakamoto, Wallberg and Yamazaki extends this viewpoint to
spacetime-dependent couplings and shows that the resulting classical
integrability conditions reproduce the corresponding renormalization
group evolution \cite{KomatsuSakamotoWallbergYamazaki2026}.  This provides a
higher-dimensional gauge-theoretic realization of the
integrability-RG relation that underlies the time-dependent models
considered here.

Taken together, these results suggest three complementary directions.
The first is specific to the models studied here: to understand
mathematically how bosonization and refermionization reorganize the qKZ
description of the interacting spin sector into the physical-rapidity
difference structure of the massive-Thirring representation.  The
second is broader: to determine how that relation fits into the
established network linking quantum affine algebras, DAHA,
Macdonald-Ruijsenaars systems, Baxter operators, $W$-algebras,
$q$-Whittaker functions, quantum $q$-Langlands and gauge theory. The third is
 to understand whether the genuinely
quantum structures found in the present construction-in particular,
the finite shifts of the physical rapidities, the resulting analytic
difference equations and the rapidity-lattice form of the exact
wavefunction-also admit a realization within the four-dimensional
Chern-Simons framework.  Such a development could provide a common
gauge-theoretic setting for the classical Lax description,
Yang-Baxter and qKZ structures, and the quantum difference equations
that arise in the time-dependent quantum field theories.

\section{Conclusion and outlook}
\label{sec:conclusion}

In this work we have constructed an exact generalized coordinate Bethe
wavefunction for the massive Thirring model with a constant four-fermion
interaction and a time-dependent bare mass.  Starting from the
time-dependent Schrodinger equation, we first considered a
rapidity-factorized one-particle ansatz without specifying the temporal
dependence of the mass.  Compatibility of this family of solutions fixes
the bare mass to have the exponential form
\begin{equation}
    m_b(t)=m_*e^{\kappa t}.
\end{equation}
This restriction is a consequence of the Bethe-compatible rapidity
factorization and should not be interpreted as a statement that the
one-particle Dirac equation cannot be solved for other mass protocols.
The resulting exponential evolution has the multiplicative form expected
from an RG trajectory of the mass perturbation and is closely related to
the separated null-coordinate dependence obtained from the
time-dependent sine-Gordon Lax construction of
Hoare, Levine and Tseytlin.

The two-particle sector determines the scattering structure of the
many-body wavefunction.  The quartic fermion interaction term generates a scalar S-matrix associated with the scattering of particles, where
\begin{equation}
S_\mu(\theta) =-\frac{\sinh(\theta/2-\ii\mu)}{\sinh(\theta/2+\ii\mu)}.
\end{equation}
Although the individual one-particle spinors depend explicitly on space
and time, this dependence does not exist in the scalar S-matrix as the interaction strength is constant. The local scattering structure is therefore the same
hyperbolic function of rapidity difference that appears in the static
massive Thirring model.

The principal modification produced by the time dependence is already visible in the one-particle periodicity condition.  In the hatted normalization used from the outset, a spatial winding acts as the pure rapidity translation
\begin{equation}
 \widehat f_\beta(x+L,t) = \widehat f_{\beta-p}(x,t),\qquad p=\kappa L.
\end{equation}
A fixed-rapidity block is therefore not periodic, and the complete state must contain the corresponding rapidity lattices
\begin{equation}
\beta_j=\widetilde\beta_j+p n_j,\qquad n_j\in\mathbb Z.
\end{equation}
The rapidity lattice is thus forced by the one-particle dynamics together with periodic boundary conditions, before any interaction-dependent recurrence is solved.  This feature sharply distinguishes the present massive problem from the massless chiral models solved using nested generalized Bethe ansatz \cite{PasnooriQKZ2026,PasnooriBKT2026}, in which the elementary chiral functions remain arbitrary and the nontrivial difference equations first arise at the level of many-body transport.

The interaction supplies the second ingredient.  In the hatted normalization the one-particle winding phase has been removed from the transport amplitude, so the scalar difference equation contains only the two-particle scalar S-matrix,
\begin{equation}
 \Phi_p(\theta-p)=S_\mu(\theta)\Phi_p(\theta).
\end{equation}
Its solution is expressed in terms of $q$-Pochhammer products or, equivalently, $q$-gamma functions with $q=e^{-p}$.  The $N$-particle reference amplitude therefore factorizes purely into pairwise functions,
\begin{equation}
A_N(\boldsymbol\beta)=\prod_{i<j}\Phi_p(\beta_i-\beta_j),
\end{equation}
while the complete periodic wavefunction is the formal sum over the independent physical rapidity lattices.  The associated scalar transport operators satisfy a discrete zero-curvature condition. For the comparison with the static model, Sec.~\ref{sec:staticlimit} introduces a plane-wave normalization in which the one-particle winding phase is explicit.  Its $\kappa\to0$ limit is $e^{\ii m_*L\sinh\beta}$, and the finite-step transport equations then reduce directly to the conventional Bethe equations.  The hatted basis is therefore the natural one for the exact time-dependent construction, while the second normalization is needed only to display the familiar Bethe quantization in the static limit.

The coordinate construction also admits a complementary momentum-space
description.  Fourier transformation of the one-particle equation
reduces the time dependence to a pair of second-order equations which,
for the exponential mass protocol, are Whittaker equations.  An explicit
integral transform relates the rapidity-dependent coordinate blocks to
the corresponding momentum-resolved Whittaker solutions.  Poisson
summation of the one-particle rapidity lattice further shows how the
discrete spatial momenta on the circle emerge from the rapidity-shift
structure.  Thus the analytic-difference and differential descriptions
are two complementary representations of the same time-dependent
one-particle dynamics.

These results also clarify the relation of the present coordinate
construction to the nested generalized Bethe ansatz developed for
time-dependent theories with internal degrees of freedom.  In the
nested problem, qKZ equations act on vector-valued amplitudes and their
off-shell Bethe-ansatz solutions contain scalar analytic difference
equations for auxiliary-rapidity weights.  In the massive Thirring
model studied here, the nontrivial recurrence acts directly on the
physical particle rapidities and is scalar from the outset.
Bosonization and refermionization relate the interacting spin sector of
the $U(1)$ Thirring model, sine-Gordon model, and the massive Thirring
model considered here.  This provides a physical setting in
which qKZ-type difference structures and Whittaker-type differential
structures arise in different representations of closely related
integrable theories.

The appearance of these structures should, however, be distinguished
from the more general mathematical qKZ-Whittaker correspondences.
Quantum affine KZ equations are related through Hecke algebra and DAHA to Macdonald-Ruijsenaars systems, and appropriate limits
lead to $q$-Toda and $q$-Whittaker functions.  The Whittaker functions
encountered here are instead the ordinary one-variable Whittaker
functions arising from the time-dependent Dirac equation.  Likewise,
the parameter $q=e^{-\kappa L}$ in the present construction measures
the physical rapidity shift generated by spatial winding and should not
be identified without further analysis with a quantum-group deformation
parameter.  The correspondence established here is therefore a
model-specific connection between two exact representations, rather
than a claim of a new general equivalence between qKZ and Whittaker
equations.

Several physical questions remain open.  The present work constructs
the exact finite-$N$ wavefunction but does not yet determine the
time evolution of a chosen many-body state or the long time dynamics. Determining how a physically
prepared state evolves, and extracting its time-dependent mass scales and correlation functions, are natural next
applications of the exact wavefunction obtained here. The repulsive regime provides a particularly simple setting for this
analysis because the static spectrum contains no breather bound states.
The attractive regime is richer, with soliton-antisoliton bound states \cite{Dashen1975,Dashen1975as},
and will require tracking how the corresponding bound-state structures
are represented in the driven spectral sum.  Open boundaries constitute
another natural extension.  The static massive Thirring/sine-Gordon
model has nontrivial boundary spectra \cite{skorik,gosh,PasnooriMT2025,PasnooriSGPT,PAA1,PAA2}, while generalized
Bethe ansatz construction for systems with time-dependent boundaries leads to boundary qKZ equations in the
nested theories \cite{pasnooriBoundary2026}. In addition, one can consider coupling the boundaries to a quantum degrees of freedom, such as an impurity \cite{Parmeshkondo1,Parmeshkondo2}. One can also consider the case of the driven boundary SG model, which in the fermionic representation would be equivalent to the bulk without the mass term that is coupled to a boundary, where the left and right moving fermions are exchanged with a time-dependent strength. Extending the present coordinate construction to the above mentioned examples would provide a direct framework for studying time-dependent boundary states. These models and their associated phenomena can be readily realized using Josephson junction arrays \cite{roy1,circuitSPT}, providing a viable platform for near-term experimental implementation.

\section*{Acknowledgments}
P. R. P is supported by Mani L. Bhaumik Prize Post Doctoral Fellowship. We acknowledge the use of ChatGPT and Google AI mode for testing our ansatz, streamlining our literature review and improving the language. All mathematical results were explicitly verified using traditional pencil-and-paper derivations. We estimate that the integration of these AI tools reduced our overall working time threefold.

\appendix

\section{Special-function identities and sign conventions}
\label{app:special}

The defining product and recurrence are
\begin{equation}
 (z;q)_\infty=\prod_{n=0}^\infty(1-zq^n),\qquad (zq;q)_\infty=\frac{(z;q)_\infty}{1-z},\qquad |q|<1.
\end{equation}
Together with the definition of $\Gamma_q$ they give
\begin{equation}
 \Gamma_q(z+1)=\frac{1-q^z}{1-q}\Gamma_q(z),
\end{equation}
and establish Eqs.~\eqref{eq:phiproduct}-\eqref{eq:phigamma} \cite{DLMF,GasperRahman2004}. To compare the spectral convention with a formula written directly in terms of $g_0$, define
\begin{equation}
 r_g=\frac{1+\ii g_0/2}{1-\ii g_0/2},\qquad \eta_g=\frac{\log r_g}{\log q}.
\end{equation}
Equation~\eqref{eq:mu} implies $r=r_g^{-1}$. Hence the identical solution can be written
\begin{equation}
 \Phi_p(\theta)=e^{\eta_g\theta}\frac{(-r_g e^\theta;q)_\infty}{(-r_g^{-1}e^\theta;q)_\infty}.
\label{eq:gproduct}
\end{equation}
An apparent interchange of numerator and denominator is therefore accompanied by $r_g\leftrightarrow r^{-1}$. Inverting only the product at a fixed parameter changes the recurrence and violates the seam condition. The explicit definition of the scalar S-matrix in Eq.~\eqref{eq:Sdef} removes this ambiguity.

\bibliographystyle{JHEP}
\bibliography{MTtime}
\end{document}